\documentclass[runningheads,openany]{svmult}

\usepackage{makeidx}   
\usepackage{graphicx}  
\usepackage{subeqnar}  
\usepackage{multicol}  
\usepackage{physprbb}    
\makeindex             

\usepackage{amsmath}
\usepackage{amsfonts}
\usepackage{theorem}

\usepackage{epsfig}

\newcommand{\con}{{\mathcal K}_d}
\newcommand{\N}{{\mathbb N}}
\newcommand{\tr}{{\mathbf t}}
\newcommand{\br}{{\mathbf r}}
\newcommand{\bv}{{\mathbf v}}
\newcommand{\trace}{{\rm Tr}}
\newcommand{\rb}{{\mathbf r}}
\newcommand{\n}{{\mathbf n}}
\newcommand{\dd}{{\rm d}}
\newcommand{\x}{{\mathbf{x}}}
\newcommand{\p}{{\mathbf{p}}}

\newcommand{\f}{{\rm f}}
\def\R{\mathbb R}

\newcommand{\mpch}{{\ifmmode{h^{-1}{\rm Mpc}}\else{$h^{-1}$Mpc}\fi}}
\newcommand{\mpc}{{\ifmmode{{\rm Mpc}}\else{Mpc}\fi}}

\newcommand{\cov}{{\ifmmode{\text{{\it cov}}}\else{ {\it cov} }\fi}}
\newcommand{\cor}{{\ifmmode{\text{{\it cor}}}\else{ {\it cor} }\fi}}
\newcommand{\var}{{\ifmmode{\text{{\it var}}}\else{ {\it var} }\fi}}

\begin{document}

\title{Vector- and tensor-valued descriptors for spatial patterns}
\author{Claus Beisbart$^{1,2}$, Robert Dahlke$^2$, Klaus Mecke$^{3,4}$, and Herbert Wagner$^2$}
\authorrunning{Beisbart et al.}
\institute{
University of Oxford, 
Nuclear \& Astrophysics Laboratory, Keble Road, Oxford OX1 3RH, U.K.
\and
Ludwig-Maximilians-Universit\"{a}t,  Sektion
Physik,  Theresienstra{\ss}e 37,  D-80333  M\"{u}nchen, Germany
\and
Max-Planck-Institut f\"ur Metallforschung, Heisenbergstr. 1, D-70569
Stuttgart, Germany   
\and
Institut f\"ur Theoretische und Angewandte Physik, Fakult\"at f\"ur
Physik, Universit\"at Stuttgart, Pfaffenwaldring 57, D-70569 Stuttgart, 
Germany
\\[5mm]  emails:
beisbart/dahlke/wagner@theorie.physik.uni-muenchen.de,\\mecke@mecke@fluids.mpi-stuttgart.mpg.de}

\date{draft \today}

\maketitle
\begin{abstract}
Higher-rank Minkowski valuations  are efficient means for describing
the  geometry and  connectivity of  spatial patterns.  We show  how to
extend the  framework of the  scalar Minkowski valuations  to vector-
and  tensor-valued measures.  The  versatility of  these measures  is
demonstrated by using simple toy models as well as real data.
\end{abstract}


\section{Introduction}
\label{sec:beisbart_in}
\begin{figure}
\begin{minipage}[t]{.99\linewidth}
\centering
\epsfig{file=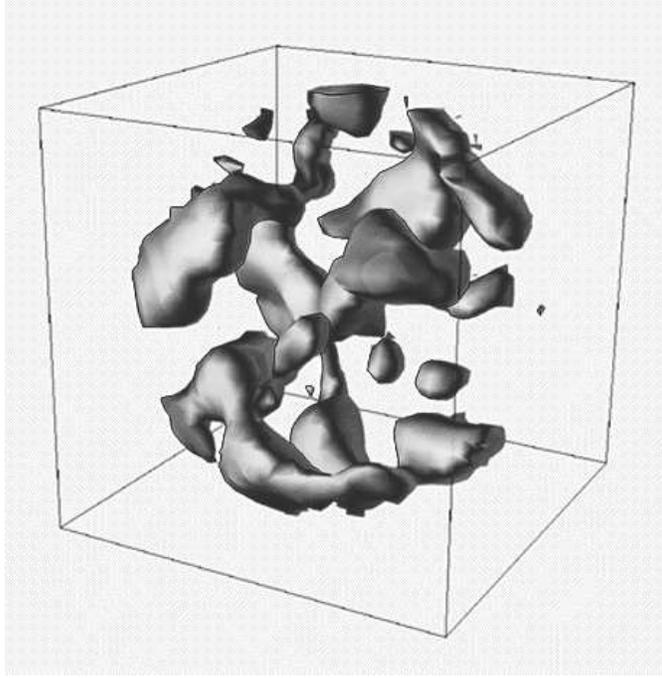,height=9cm}
\end{minipage}\hfill
\caption{ The matter  distribution around our galaxy as  traced by the
PSCz survey.  The galaxy number  density is smoothed;  one isodensity
contour     is     shown     (from     the    PSCz     homepage     at
http://star-www.dur.ac.uk/cosmology/theory/pscz.html).
\label{fig:beisbart_pscz}}
\end{figure}

The spatial  patterns originating from the  polymorphic aggregation of
matter  in Nature  occur on  vastly different  length scales  and with
unlimited   variety.     For   instance,   the    pattern   shown   in
Figure~\ref{fig:beisbart_pscz} could represent  a biological structure obtained
with X-ray tomography, or a  colloidal cluster in a milky emulsion. In
fact, Figure~\ref{fig:beisbart_pscz}  displays an isodensity  contour of cosmic
matter as traced by the galaxies surrounding the Milky way, which lies
in the center of the plot.
\\
The visual  impression from such pictures may  already convey valuable
insight into the dynamic evolution or -- in the context of biology -- into
the functionality of such  structures. However, for unbiased inferences
as  well as  for the  comparison  with model  simulations, one  needs
objective and  quantitative measures to characterize  the geometry and
topology of typical structural motifs.
%
\\
%
In this  article we deal  with a family  of functions which  provide a
unique description of the morphology displayed by spatial patterns. We
call these functions the  Minkowski valuations; they are extensions of
the  well-known  Minkowski  functionals  often mentioned  within  this
volume (see  the contributions by  C.~Arns et al. and  M.~L\"osche and
P.~Kr\"uger in this volume as well as \cite{beisbart_mecke:additivity}
in the  previous volume  LNP~554).  Therefore they  share some  of the
simplicity  and  beauty  which  the  Minkowski  functionals  own.   In
particular,  they  are  characterized   by  simple  axioms  which  set
intuitively reasonable standards  for a morphological description.  In
the same way as the Minkowski functionals generalize the notion of the
volume, the Minkowski valuations extend  the notion of the mass vector
(or  center of mass)  and of  the inertia  tensor.  Like  their scalar
relatives  the  majority of  the  Minkowski  valuations are  curvature
measures integrated over  the surface of a pattern.\\  We start with a
short  motivation in  Section~\ref{sec:beisbart_mot}  by showing  that
there are good reasons to  go beyond the scalar Minkowski functionals.
After  having outlined  ways to  generalize the  Minkowski functionals
(Section~\ref{sec:beisbart_hi}),  we review  some  of the  extensions'
mathematical   properties   in   Section~\ref{sec:beisbart_exp}.    To
illustrate  how  useful  the  Minkowski  valuations  are,  we  discuss
concrete applications covering  submolecular up to cosmological scales
(Sections~\ref{sec:beisbart_ap}     and~\ref{sec:beisbart_an}).     \\
Before,  two cautionary remarks  are appropriate:  first, in  order to
distinguish conceptually between  the scalar Minkowski functionals and
their higher-rank  extensions, we shall address the  whole entirety of
Minkowski functionals  and their  extensions via the  term ``Minkowski
valuations''  (MVs),   whereas  ``Minkowski  functionals''   (MFs)  is
reserved  to the  scalar Minkowski  functionals as  before.  Secondly,
there  is a  vivid  theoretical development  of higher-rank  Minkowski
valuations                          these                         days
{}\cite{beisbart_schneider:tensor,beisbart_schneider:tensor2,beisbart_beisbart:tensor}.
Therefore,  this  overview  can  not  provide  a  complete  survey  on
Minkowski valuations;  on the contrary, substantial progress  is to be
expected both from the mathematical side as well as from applications.
For instance, morphological analysis based on Minkowski valuations may
be useful to quantify molecular orientations (see the contributions by
F.~Schmid and  N.~H.~Phuong, U.~S.~Schwarz and  G.~Gompper, D.~Jeulin,
C.~Arns  et al., and  H.-J.~Vogel in  this volume).  Our vector-valued
descriptors may also  be used to refine the  morphological analysis of
Langmuir monolayer phases (presented by M.~L\"osche and P.~Kr\"uger in
this volume) which has so far been based merely on scalar MFs. Another
application may  cover the geometric characterization  of liquid foams
described by F.~Graner in this volume, where stress tensors need to be
related to a complex spatial structure.
\section{Going beyond scalar Minkowski functionals -- a physical motivation}
\label{sec:beisbart_mot}
\begin{figure}
\begin{center}
\epsfig{file=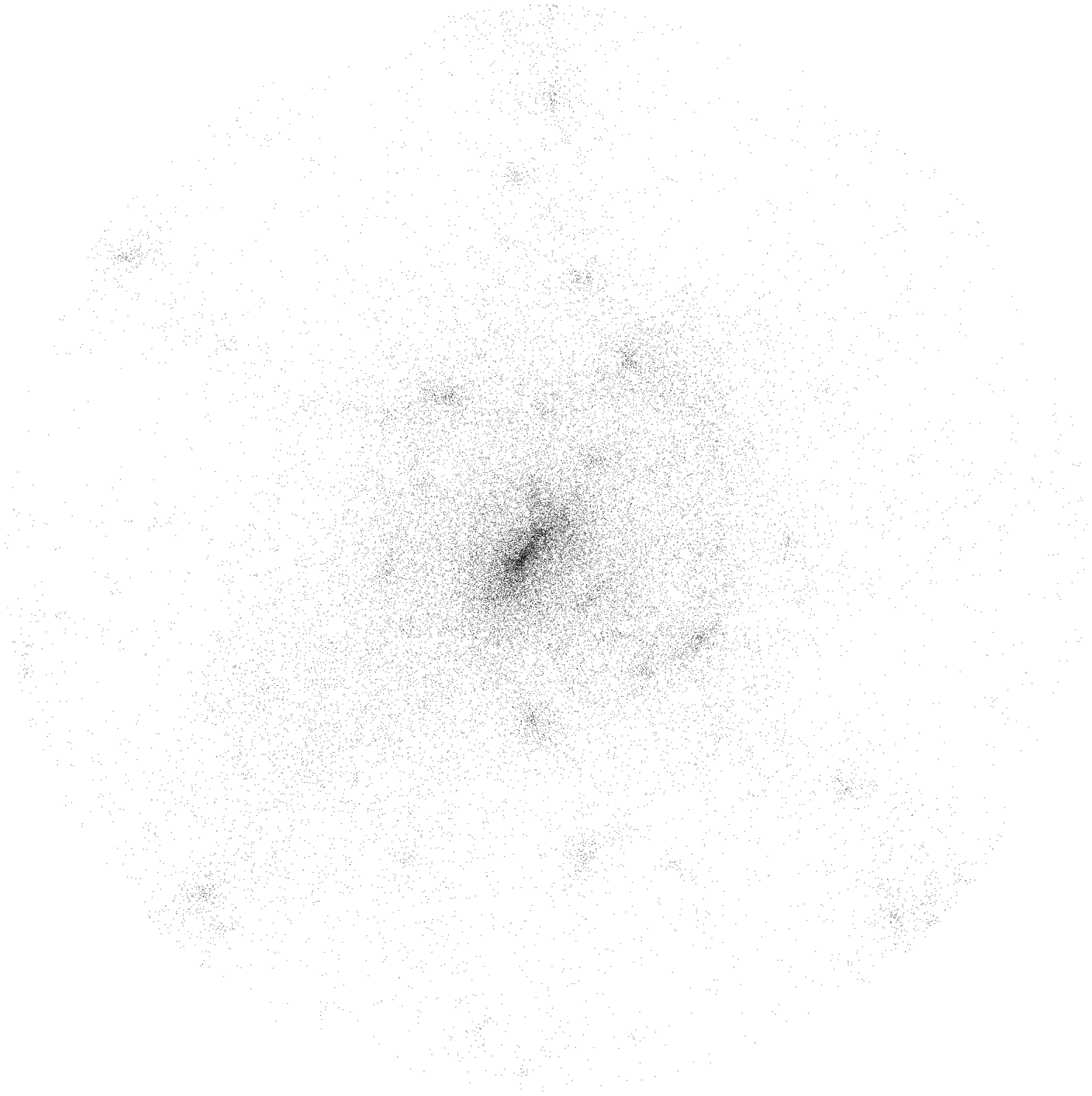,height=5cm}
\epsfig{file=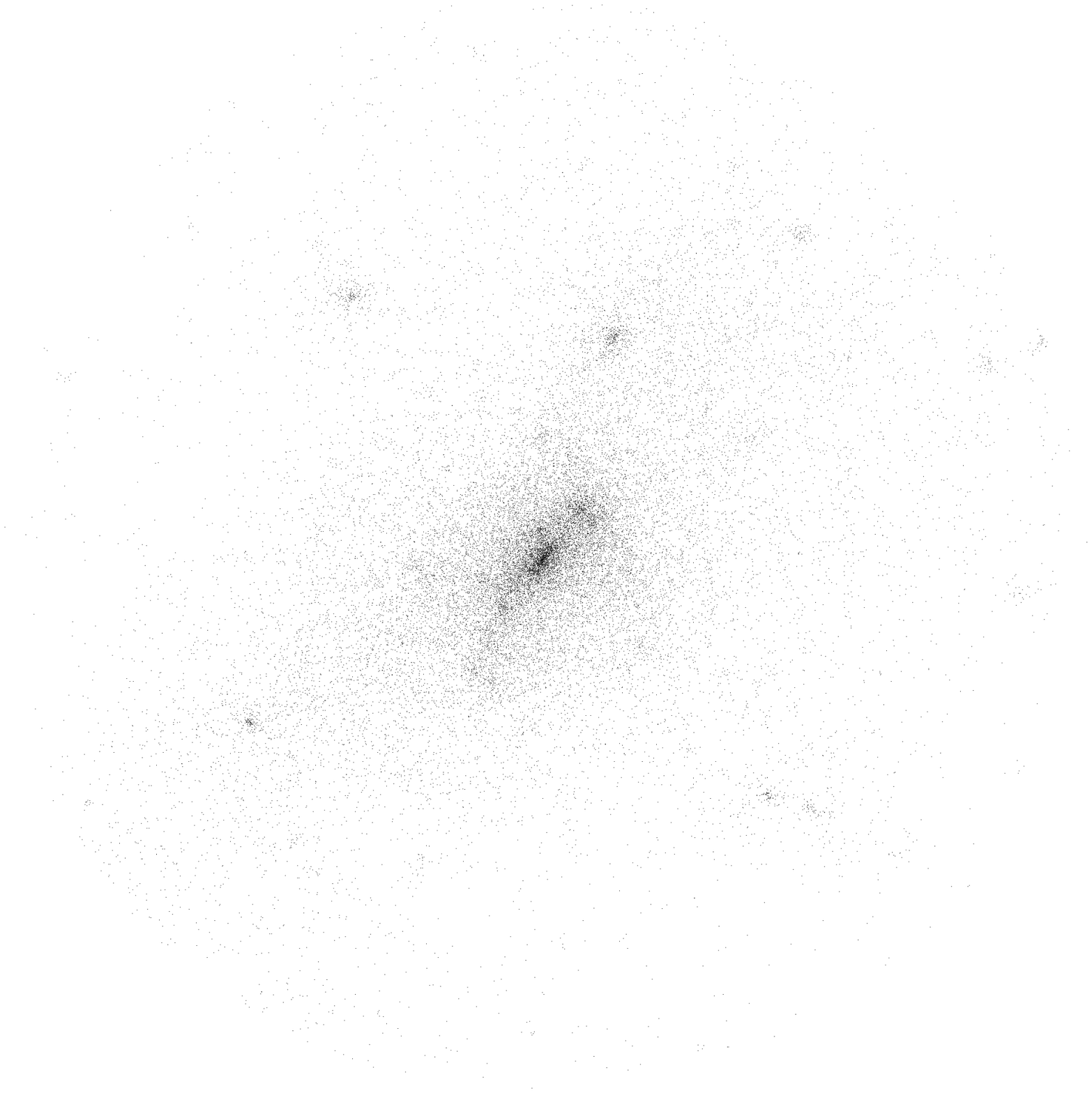,height=5cm}
\end{center}
\caption{The  projected matter  distribution of  two  simulated galaxy
clusters (from the GIF-project). Their different morphologies are due
to   different  values  of  the cosmological  parameters  and  to
different initial power spectra  imposed (left panel: $\tau$ Cold Dark
Matter  model;  right  panel  open   Cold  Dark  Matter  model,  see
{}\cite{beisbart_bartelmann:arcIV}  for   more  details).   Apart   from  that,
however, the  clusters stem from comparable initial conditions. In  order to
distinguish  between  both models  we  need  a thorough  morphological
description.\label{fig:beisbart_problem} }
\end{figure}

\begin{figure}
\begin{center}
\epsfig{file=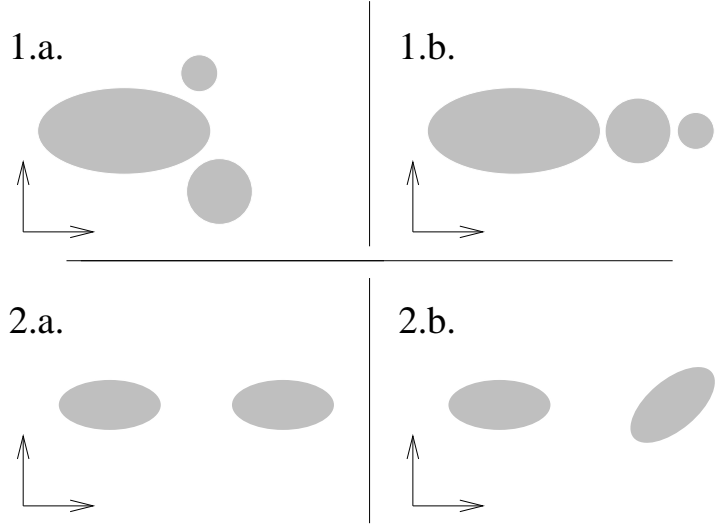,height=7cm}
\end{center}
\caption{Neither the  schematic clusters  displayed in panel  1.a. and
1.b. nor  both patterns  in panel 2.a.  and 2.b. can  be distinguished
from    each     other    using    only     the    scalar    Minkowski
functionals.  Additivity and motion-invariance  prevent the  MFs from
discriminating  between  the  different  positions (upper  panels)  or
different   orientations   (lower   panels)  of   the   subcomponents,
respectively.\label{fig:beisbart_need}}
\end{figure}

The following problem is  typical for challenges occurring in physics:
the formation of  galaxy clusters depends in a  complicated way on the
background  cosmology usually  described  in terms  of a  cosmological
model.  Such a model is defined through the values of the cosmological
parameters   and   the  power   spectrum   of   the  initial   density
perturbations {}\cite{beisbart_peebles:principles,beisbart_bartelmann:arcIV}.   In  order
to  investigate  the impact  of  the  background  cosmology on  galaxy
clusters one carries  out gravitational $N$-body simulations; typical
results   comprise   cluster   images   such   as   those   shown   in
Figure~\ref{fig:beisbart_problem}.  To  compare the  models and to  check which
model fits  best present-day  cluster observations, one  has to use
quantitative and discriminative morphological measures.
\\
The  Minkowski  functionals certainly  place  practical  tools at  the
physicist's disposal.   Defined on the  convex ring $\con$,  i.e.  for
finite  unions of  compact, convex  sets in  $d$-dimensional Euclidean
space,  they  are the  only  motion-invariant, convex-continuous,  and
additive  descriptors.   Despite  this very  general  characterization
there  are  only a  finite  number  of  Minkowski functionals,  namely
$(d+1)$  in  $d$  dimensions  (characterization  theorem,
\cite{beisbart_hadwiger:vorlesung,beisbart_klain:hadwiger}).       They
have simple geometrical meanings  derived from their representation as
integrals.  These  integrals either extend over the  pattern itself to
yield  the  volume,  or  cover  its surface  which  is  weighted  with
combinations  of the  local  curvatures in  order  to give  -- in  two
dimensions, e.g.  -- the perimeter and the integrated curvature (which
equals  the  Euler characteristic).   The  Minkowski functionals  also
arise quite naturally as expansion coefficients in the Steiner formula
{}\cite{beisbart_schneider:brunn}   specifying  the   volume   of  the
parallel body of a convex body. Indeed, the Minkowski functionals have
been extensively  applied to  physical structures at  vastly different
length  scales, see  \cite{beisbart_mecke:additivity} for  an overview
and     \cite{beisbart_kerscher:statistical}     for     astrophysical
applications.  On  the mathematical side, numerous  useful results such
as  the   principal  kinematic   formulae  or  the   Crofton  formulae
(see {}\cite{beisbart_hadwiger:vorlesung})  have been  proven  with integral
geometric   methods.\\  However,   the   morphometry  with   Minkowski
functionals  is by  no means  comprehensive.  Combining  two  of their
defining principles,  additivity and homogeneity, one  can easily see,
that,  e.g.,  moving  a  connected   part  of  a  pattern  results  in
significant visible changes which can not be detected by the Minkowski
functionals  at  all.  This  is  illustrated  in  the first  panel  of
Figure~\ref{fig:beisbart_need} representing schematic clusters.  It is
surely of  interest, however, in which  way the subclumps  of a system
are arranged, and how they are aligned with respect to each other (see the
second panel  of Figure~\ref{fig:beisbart_need}).  Therefore,  we have
to move  beyond the scalar  Minkowski functionals.  It turns  out that
one  can  generalize the  Minkowski  functionals  without leaving  the
mathematical framework of integral geometry.
\section{The hierarchy of Minkowski valuations -- extending the framework}
\label{sec:beisbart_hi}
There are several ways to extend the framework of the Minkowski
functionals. For instance, one may start with Steiner's formula.
\paragraph{The Steiner formula.} 
It is a natural choice to  employ the volume $V (K)=\int_{K} \dd V$ of
a  body $K$ for  its very  coarse description.   If the  parallel body
$K_\epsilon$ is  considered, which contains  all points closer  to $K$
than $\epsilon>0$, and if $K$ is convex, then one finds from Steiner's
formula an expansion for $V(K_\epsilon)$ in powers of $\epsilon$,
\begin{equation}
V\left(K_\epsilon\right) = \sum_{\nu=0}^{d}\binom{d}{\nu}\epsilon^\nu W_{\nu}\;\;\;,
\end{equation}
where  the  Minkowski   functionals  $W_{\nu}$  arise  as  expansion
coefficients. The latter functionals are useful morphometric concepts
and can be extended to non-convex patterns. In this sense, the volume
quite generally leads to a family of related measures.
\\
Since we  saw that  in some applications  (recall the upper  panels of
Figure~\ref{fig:beisbart_need})   the  position   of  a   body  may   be  worth
considering, we specify the location of $K$ with its {\em mass vector}
\begin{equation}
V^1 \left(K\right) \equiv \int_K \dd V \x \;\;\;.\label{eq:beisbart_def_massvector}
\end{equation}
Alternatively,  we  could introduce its center  of  mass  $\p_0\equiv
V^1/V$. 
\\
Moreover,  as the  lower panels  of Figure~\ref{fig:beisbart_need} indicate,
the body's  orientation may be relevant. In this case we have to move on to even higher moments; for instance, the tensor
\begin{equation}
V^2 \left(K\right) \equiv \int_K \dd V \x \x \label{eq:beisbart_def_inertia}
\end{equation}
(where  $\x\x$   denotes  the  symmetric   tensor  product\footnote{As
\cite{beisbart_beisbart:tensor} show,  there are no  non-trivial {\em anti}symmetric
Minkowski   valuations   of   rank   $r\geq1$   at   all,   i.e.    no
motion-covariant,   additive,  and  conditionally   continuous  tensor
functions with rank  $r\geq1$, see below Section~\ref{sec:beisbart_hi} for the explanation of these
concepts.}) contains information on  main directions of the body $K$.
Note, that both the mass vector  of a body and its inertia tensor (the
traceless  part  of Equation  \ref{eq:beisbart_def_inertia}:  $V^2(K) -  \trace
\left(V^2(K)\right)   E_d$,  where   $E_d$   is  the   $d$-dimensional
second-rank unit tensor) are of  major interest for the physicist: the
mass  center of a  rigid body  moves in  space as  if the  whole force
acting  on the  body was  concentrated onto  a point  particle  at its
center of mass.  The inertia tensor determines how a body is rotating,
given  a fixed  angular  momentum.  --  Higher-rank  moments such  as
symmetric  $r$-rank tensors  of the  form $V^r  \left(K\right) \equiv
\int \dd V \x^r$ with $r>2$, may be considered, too.
\\
Now it is again possible to write down a Steiner-type formula for the
transition           to           the          parallel           body
of a convex set $K$
{}\cite{beisbart_schneider:schwerpunkte,beisbart_schneider:tensor}: the
scalar  expansion coefficients  have  to be  replaced  by vector-  or
tensor-valued coefficients. The following simple argument shows that
not only  the character,  but also the  number of coefficients  has to
change:  for  a  unit  ball  centered at  the  origin,  $B_R(0)\equiv
\big\{\x\big\vert |\x |\leq R\big\}$, the construction of the parallel
body  results in  $B_{(\epsilon+R)}(0)$.   On the  other  hand, as  a
simple               calculation              shows,              $V^r
\left(B_{(\epsilon+R)}(\x)\right)\propto
\left(\epsilon+R\right)^{d+r}$.  Thus,  powers up to  $(d+r)$ arise in
the  generalization of  Steiner's  formula,   which   we
write formally as
\begin{equation}
V^r\left(K_\epsilon\right) = \sum_{\nu =0}^{d+r}\epsilon^\nu M_{r,\nu}\;\;\;.
\end{equation}
As for the scalar Minkowski functionals,  
one can  obtain an  integral representation  of  the Steiner
coefficients by  considering a  smooth and convex  body $K$.  
\begin{figure}[h]
\begin{center}
\epsfig{file=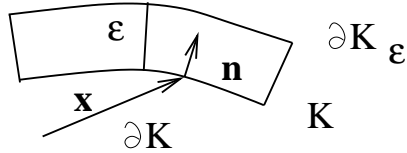,height=2cm}
\end{center}
\caption{The fact that each volume element in $K$ is weighted with the
position vector in  order to give $V^r(K)$ for  $r>0$, makes the local
normal $\n$ relevant for the parallel body $K_\epsilon$.
\label{fig:beisbart_closer}}
\end{figure}
From the
integral representation  of the  scalar MFs and  from the form  of the
moments $V^r$  one might guess  that the coefficients  $M_{r,\nu}$ are
built by functions of the form:
\begin{equation}
\int_{\partial K}\dd S^{d-1} \x^r s_\nu \left(\kappa_1,..,\kappa_{d-1}\right)
\end{equation}
with the  elementary symmetric functions  $s_\nu$ for $\nu=0,..,(d-1)$
of the local  curvatures $\kappa_1,..,\kappa_{d-1}$ (for a definition,
see~\cite{beisbart_schneider:tensor}). But  this will not  suffice, since we
need    $(d+r)$   coefficients.     A   closer    look    shows   (see
Figure~\ref{fig:beisbart_closer}),  that indeed the  local normal  vector comes
into  play; for, if  $\x\subset \partial  K$, then  $\x+\epsilon \n$
lies  on the  surface of  $K_\epsilon$;  here $\n$  denotes the  local
normal vector to  $\partial K$ at the point  $\x$.  The weighting with
the  local position  in the  definition  of $V^r$  therefore leads  to
contributions  from  local  normal  vectors in  the  Steiner  formula.
In \cite{beisbart_schneider:tensor}  it is shown   that   in  general   the   expansion
coefficients $M_{r,\nu}$ can be  written as linear combinations of the
quantities:
\begin{equation}
W_\nu^{p,q}  \equiv \frac{1}{\nu\binom{d}{\nu}}\int_{\partial K} \dd S^{d-1}s_{\nu-1} \left(\kappa_1,..,\kappa_{d-1}\right) \x^p  \n^q \label{eq:beisbart_alesker}
\end{equation}
for  $\nu=1,..,d$,  and $p,q\in  \N_0$.   Now  the $W_\nu^{p,q}$  look
  more  basic than  the  coefficients  arising  in the  Steiner
formula for the volume tensor.  It seems therefore, as if, in contrast
to the case of the scalars, the Steiner coefficients are no longer the
basic quantities.  In order to generalize the Minkowski functionals we
therefore have to move to an alternative route.
\paragraph{Spatial moments of the Minkowski functionals.} 
Let us  take the moments  $V^r$ together with  the $W_\nu^{r,s}$
($\nu=0,..,(d-1)$, $r,s=0,1,..$) as basic quantities.  In order to
arrive at
a coherent notation,  we set $W^{r,0}_0\equiv V^r$
and  $W^{r,j}_0=0$  for  $j>0$.  Higher-rank  Minkowski  functionals
therefore are higher moments of the scalar Minkowski functionals where
the local surface  is weighted with both position  and surface normal.
If the body $K$  is shifted with a translation vector $\tr$,
then the $W_\nu^{r,s}$ transform like
\begin{equation}
W_\nu^{r,s} \left( K+\tr\right) = \sum_{p=0}^{r} W_\nu^{p,j} \tr^{i-p} \binom{i}{p}\;\;\;,
\end{equation}
as follows  immediately from their definition. For rotations $R$,
  we have 
\begin{equation}
W_\nu^{r,s} \left( R K\right) =  R W_\nu^{r,s} \left( K\right)\;\;\;,
\end{equation}
where we  use the fact that the $d$-dimensional orthogonal group $O(d)$ has a natural representation
 in each tensor  space. 
\\ 
Recalling   that   the   Minkowski   functionals  can   be   described
axiomatically,  one would  be inclined  at this  point to  look  for a
characterization theorem  for vectors and  tensors. It turns  out that
one  has to  include even  more valuations  in order  to  achieve that
purpose.
\paragraph{Axiomatic characterization.} 
A rigorous  way to extend the Minkowski  functionals is therefore
to relax one  of their defining axioms.  We  keep the {\em additivity}
\begin{equation}
\Phi (A\cup B) = \Phi (A) +  \Phi (B) -  \Phi (A\cap B)
\end{equation}
for a tensor  valuation $\Phi$; we also maintain  the {\em conditional
continuity},   i.e.    continuity   for   the   subclass   of   convex
bodies\footnote{On the  space of convex bodies one  uses the Hausdorff
metric, see \cite{beisbart_schneider:brunn}, p.~48 for a definition.};
but, in  the spirit  of our previous  findings, we replace  the motion
invariance  by  {\em  motion   covariance}  defined  as  follows:  for
translations we require  the existence of a number  $n\in \N_0$ and of
functions $\Phi_i$ ($i=0,..,n$) such that for all $K$
\begin{equation}
\Phi (K+\tr) = \sum_{i=0}^n \Phi_i (K) \tr^{i}\;\;\;.
\end{equation}
Note, that products  such  as  $\Phi_i  (K)  \tr^{i}$  always  denote
symmetric  tensor   products.   For  rotations  we   use  the  natural
representation of  $O(d)$ in the space of tensors as  before, and
require for $R\in O(d)$:
\begin{equation}
\Phi (RK) = R \Phi (K)\;\;\;.
\end{equation}
In a seminal work, Alesker \cite{beisbart_alesker:tensor} proved the following
{\em     characterization     theorem}:    every     motion-covariant,
convex-continuous,  and additive mapping  $\Phi$ can  be written  as a
linear combination of the functionals
\begin{equation}
E_d^k W_{\nu}^{r,s}\label{eq:beisbart_alesker2}
\end{equation}
with   $i,j,k\in   \N_0$   and   $\nu=0,..,d$.   $E_d$   denotes   the
$d$-dimensional unit matrix\footnote{For higher than two-dimensional
spaces, it does  not make a difference whether  we require $O(d)$- or
$SO(d)$-covariance.   In  the   two-dimensional  case,  however,  it
does~\cite{beisbart_alesker:tensor}.    For  the  one-dimensional   case  see
also~\cite{beisbart_alesker:rotation}.}.  \\ 
Equation~\eqref{eq:beisbart_alesker2} gives all of the Minkowski valuations
and finally answers the question how to generalize the MFs in a
suitable manner. It contains more valuations than those given in Equation~\eqref{eq:beisbart_alesker}. 
However, some caution has to
be  exercised: Alesker's  theorem \cite{beisbart_alesker:tensor} only  bounds the  dimension  of the
tensor functional spaces from above.   It is not obvious that all
of   the   above   functionals   are  linear   independent.    Indeed,
\cite{beisbart_mcmullen:tensor} proved some linear relationships holding among
the tensors.  In  order to give one example, Gau{\ss}' theorem
yields
\begin{equation}
W_1^{1,1}=\int_{\partial K} \dd S^{d-1} \x \n = W_0^{0,0} (K) E_d\;\;\;.\label{eq:beisbart_linear}
\end{equation}
One therefore has to explore the different tensor ranks further.
\section{Exploring higher-rank Minkowski valuations}
\label{sec:beisbart_exp}
\subsection{Some mathematical results}
\paragraph{Minkowski vectors.} 
Minkowski      vectors, i.e. Minkowski valuations of rank one,     have      been      known     for      some
years~\cite{beisbart_hadwiger:vekt,beisbart_schneider:schwerpunkte,beisbart_schneider:schwerpunkte2,beisbart_hadwiger:vect2}.
There is a strict  one-to-one correspondence between the vectors and
the   scalars:    for   each   Minkowski    functional   $W_\nu\propto
\int_{\partial    K}   s_{\nu-1}\left(\kappa_1,..,\kappa_{d-1}\right)$
there    is     a   related    vector     $\int_{\partial    K}
s_{\nu-1}\left(\kappa_1,..,\kappa_{d-1}\right)\x$,  such  that $(d+1)$
vectors exist  altogether.  The  reason why no  normals enter  at this
level is  basically because for compact  convex bodies $\int_{\partial
K}       s_{\nu-1}\left(\kappa_1,..,\kappa_{d-1}\right)\n=0$\      for
$\nu=0,..,(d-1)$        (for         an        elementary        proof
see~\cite{beisbart_hadwiger:vect2}).  The  integral geometry of  vectors was
extensively   scrutinized,   see,   e.g.,~\cite{beisbart_hadwiger:vekt}.    For
physical              applications              of             vectors
(see~\cite{beisbart_beisbart:querletter,beisbart_beisbart:fp})      the     curvature
centroids
\begin{equation}
\p_i\equiv W_\nu^{1,0}/ W_\nu^{0,0}\quad {\rm (}\nu=0,..,d{\rm )}
\end{equation}
are a useful concept.  Their meaning is straight forward: they locate
morphological information.  We shall  see below that, if the curvature
centroids  do  not coincide  within  one  point,  strong evidence  for
asymmetry is given.
\paragraph{Second-rank tensors.} The situation becomes more complicated for
higher-rank  tensors. It  can be  shown,  that at  the second  rank,
the following 
$(3d+1)$ tensors provide a base~\cite{beisbart_beisbart:tensor}:
\begin{alignat}{1}
&\int_K \dd V \x \x\;\;\;, \notag\\
&\int_{\partial K} \dd S^{d-1} s_\nu \x \x\;\;\;,
\quad  \int_{\partial K} \dd S^{d-1} s_\nu \x\n\;\;\;, \quad 
\int_{\partial K} \dd S^{d-1} s_\nu \n\n\label{eq:beisbart_second}
\end{alignat}
for $\nu=0,..,(d-1)$. Not all of  them are interesting from a physical
point of view, since some of them are related to Minkowski functionals
times the  unit tensor  and thus do  not contain any  new information.
Principal  kinematic  formulae  and  Crofton formulae  are  known  for
second-rank
tensors~\cite{beisbart_schneider:tensor,beisbart_beisbart:tensor,beisbart_schneider:tensor2}.
\\ For a geometrical interpretation, we have to bear in mind, that the
values of the tensors depend on the origin adopted.  For $W_\nu^{2,0}$
($\nu=0,..,d$), however, the corresponding curvature centroid $\p_\nu$
is   a  natural  reference   point;  the   other  tensors   listed  in
Equation~\eqref{eq:beisbart_second} are translation-invariant.  Obviously, with
such a choice the tensors are sensitive to morphological anisotropies,
in the case of symmetry  they align along the symmetry axes, otherwise
they  distinguish  morphologically   interesting  directions.   It  is
therefore useful to consider their eigenvalues and eigendirections.
\paragraph{Beyond second rank.}
Although higher-rank tensors may be interesting as well, we shall
not consider them further at this point.
\subsection{Examples}
\begin{figure} \centering
\begin{minipage}[t]{.33\linewidth}
\centering\epsfig{file=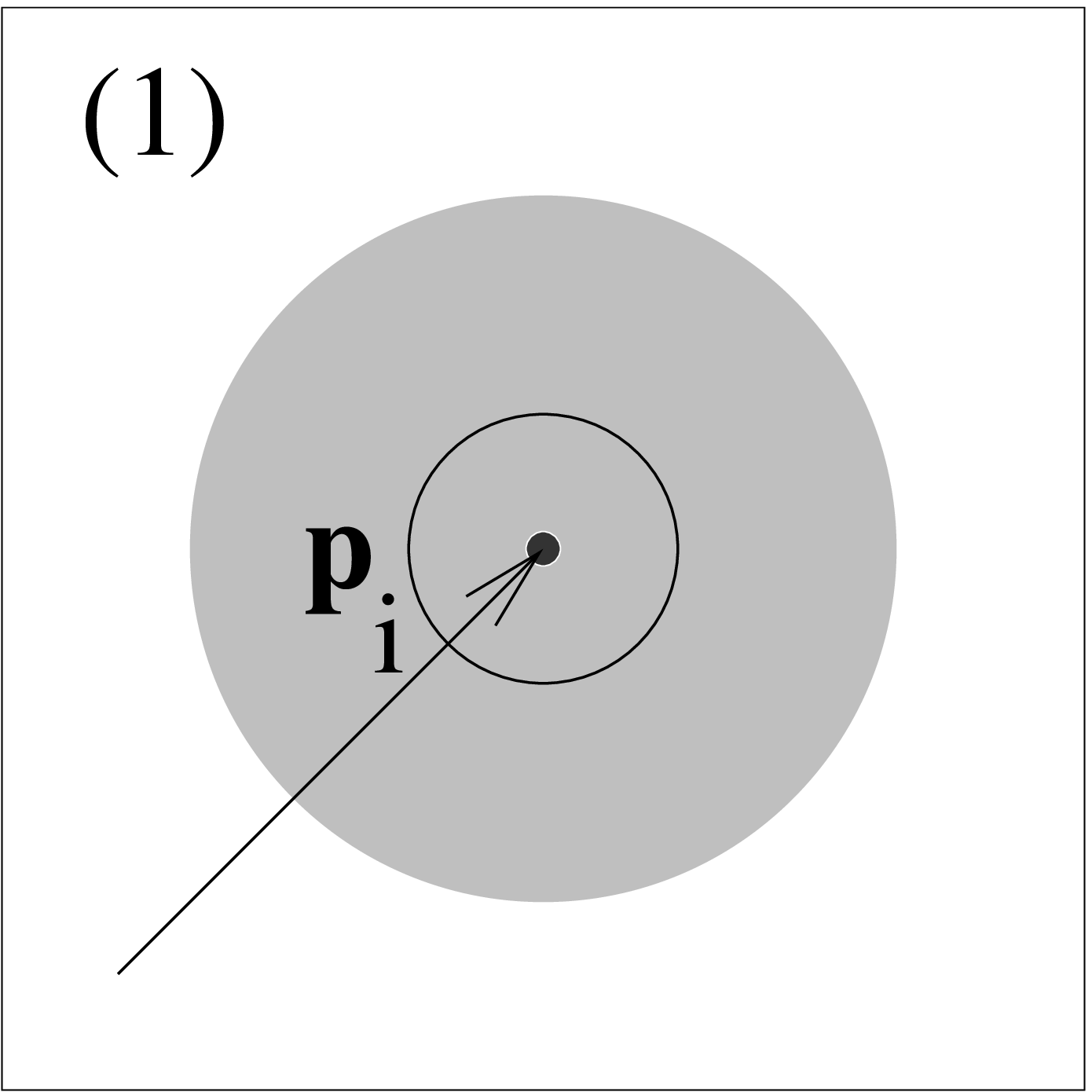,width=3.7cm}
\end{minipage}\hfill
\begin{minipage}[t]{.33\linewidth}
\centering\epsfig{file=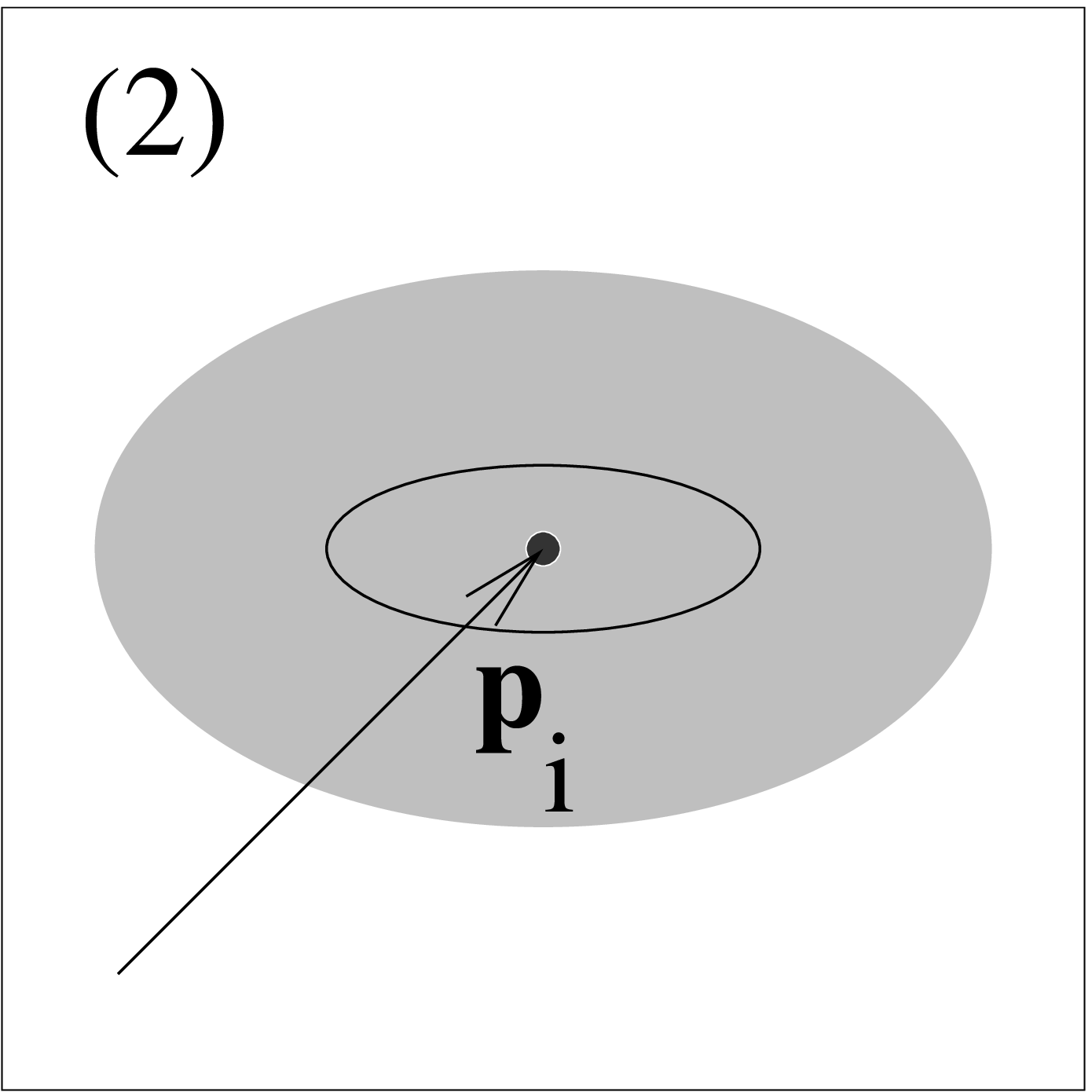,width=3.7cm}
\end{minipage}\hfill
\begin{minipage}[t]{.33\linewidth}
\centering\epsfig{file=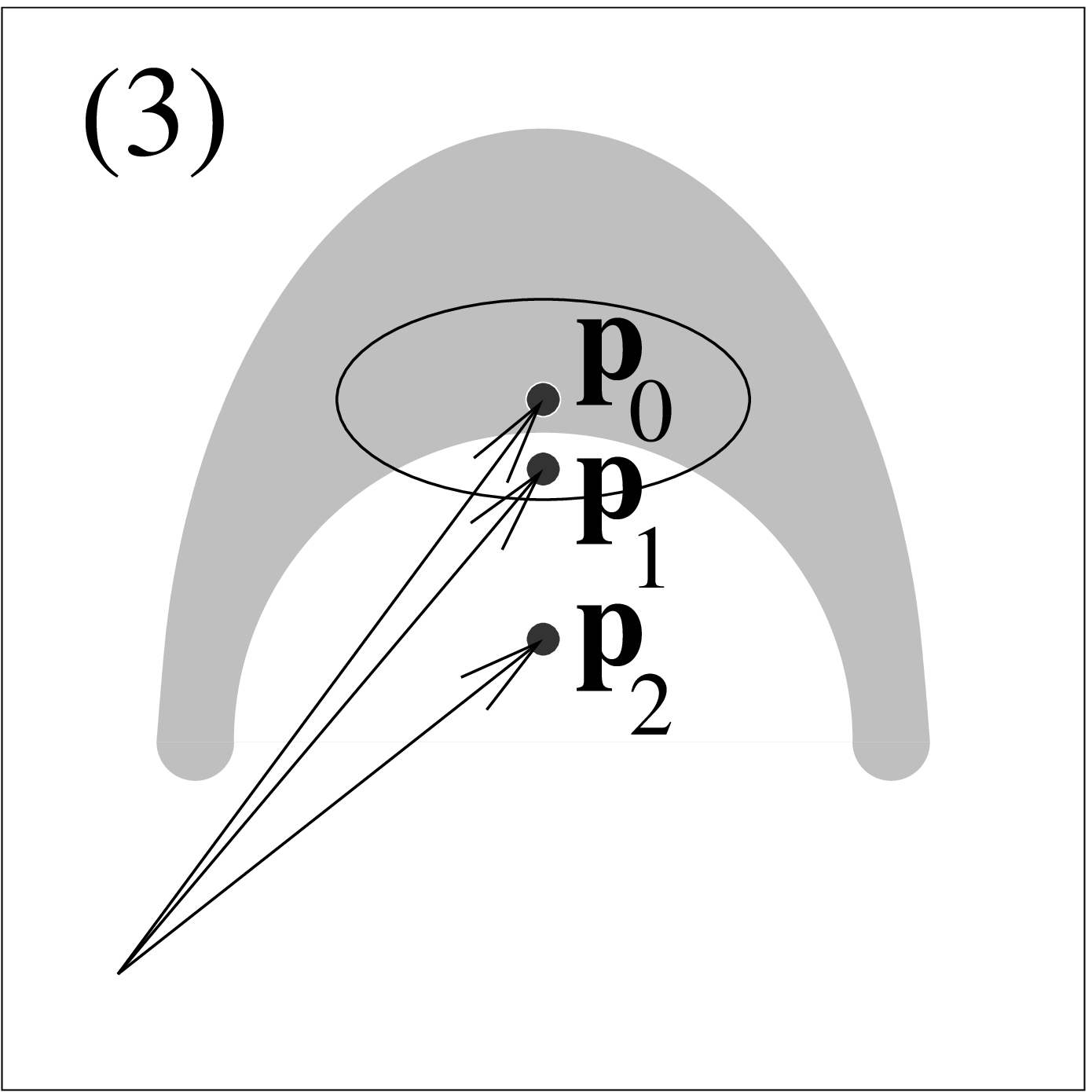,width=3.7cm}
\end{minipage}\hfill
\vspace{5mm}
\begin{minipage}[t]{.33\linewidth}
\centering\epsfig{file=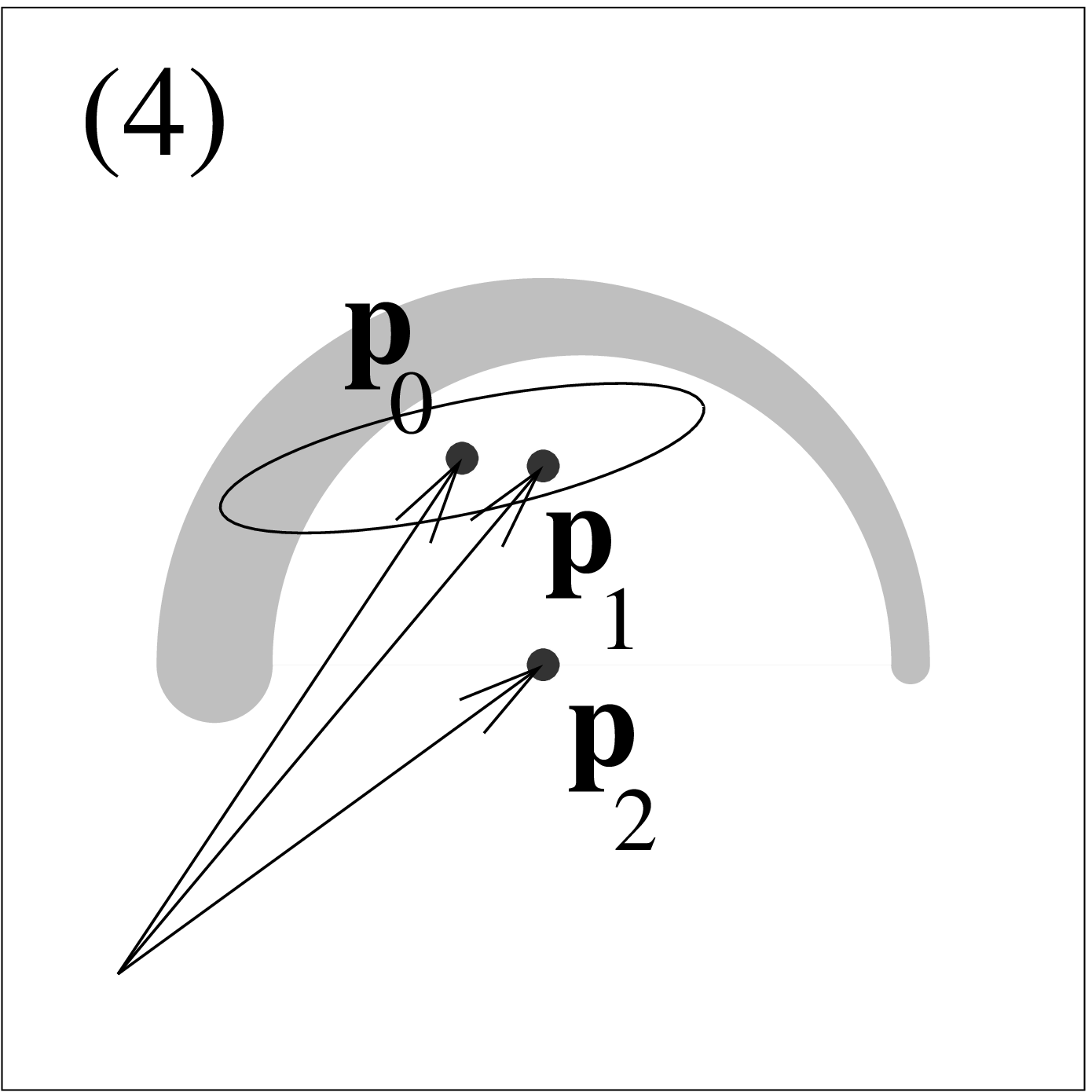,width=3.7cm}
\end{minipage}\hfill
\begin{minipage}[t]{.33\linewidth}
\centering\epsfig{file=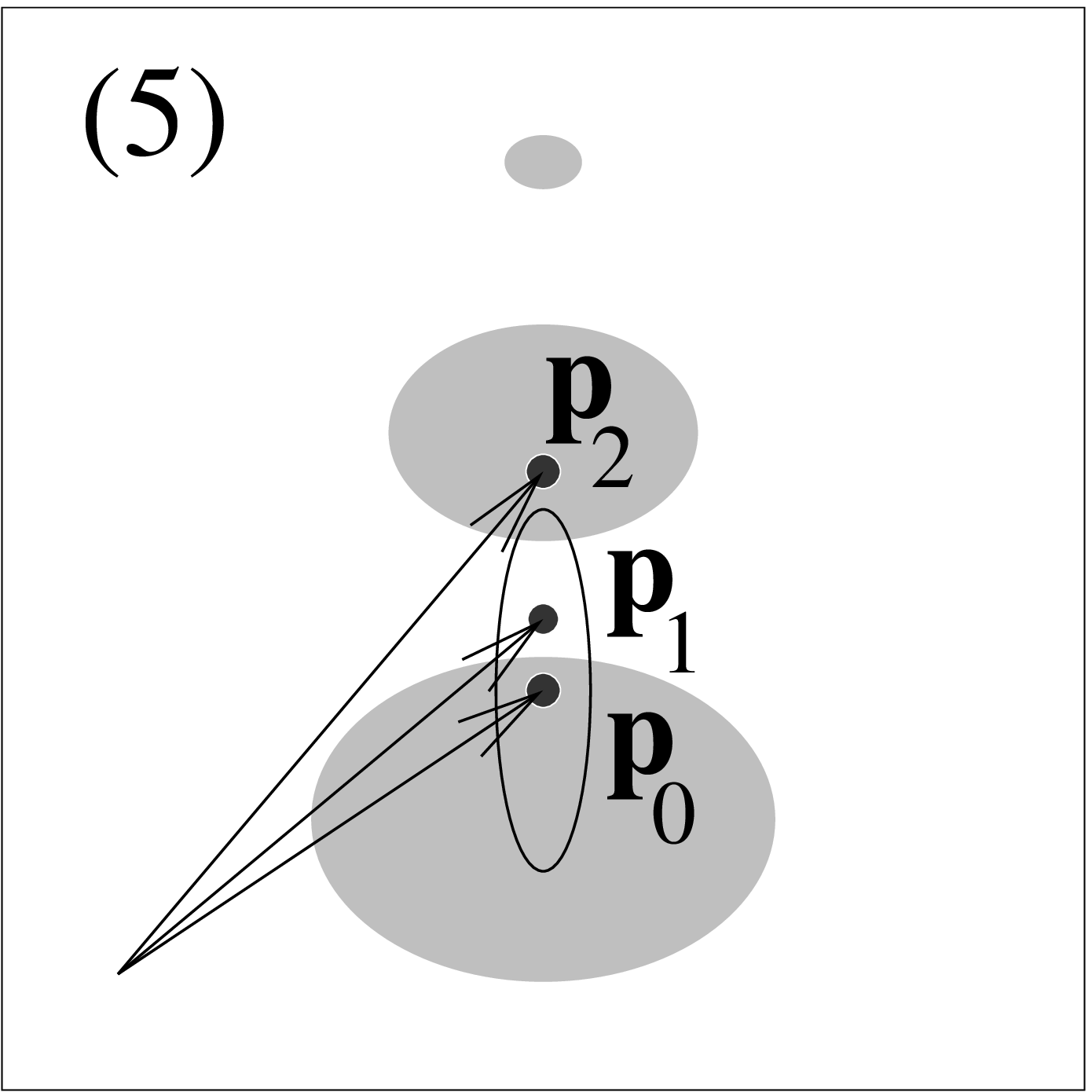,width=3.7cm}
\end{minipage}\hfill
\begin{minipage}[t]{.33\linewidth}
\centering\epsfig{file=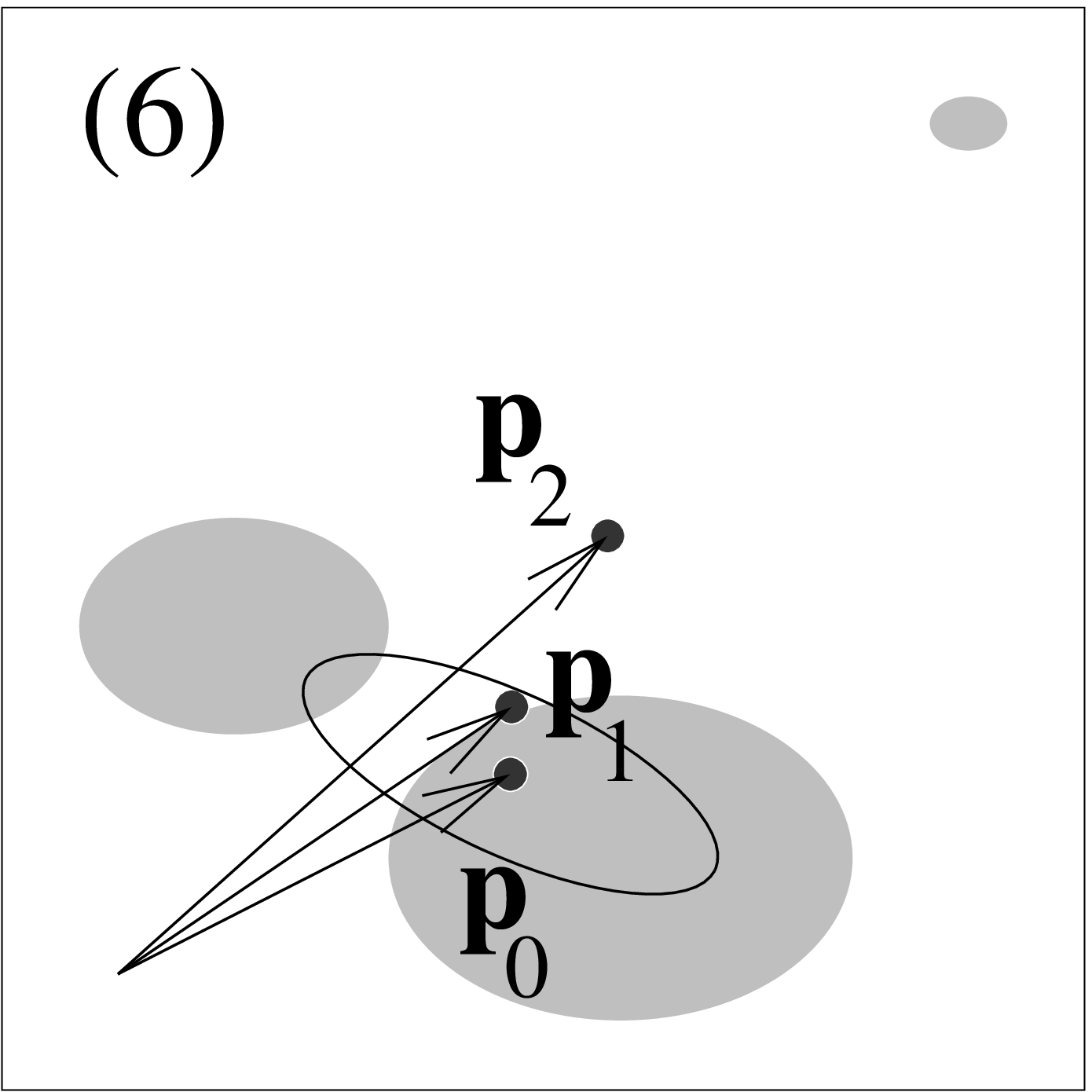,width=3.7cm}
\end{minipage}\hfill
\caption{A  number of  simple patterns  together with  their curvature
centroids $\p_i$ ($i=0,..,2$). The  ellipse represents the mass tensor
$W_0^{2,0}/W_0^{0,0}$;   it  is  calculated  with  the   center  of
mass $\p_0$ as reference point. The ellipses indicate the ratio of the
eigenvalues and the eigendirections as well.\label{fig:beisbart_examples} }
\end{figure}

\begin{figure}
\begin{center}
\centering\epsfig{file=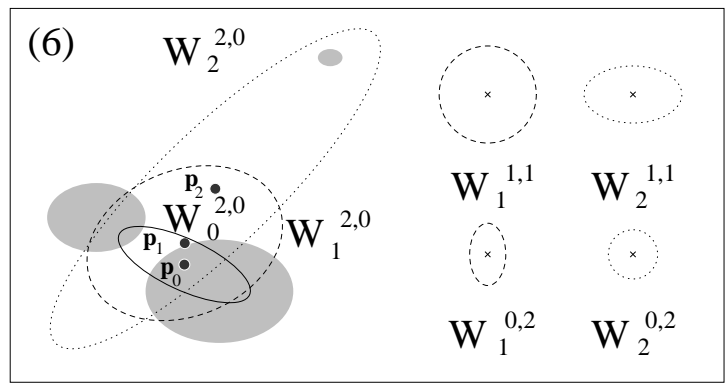,height=5cm}
\end{center}
\caption{We  show  all  of  the  tensors  $W_0^{2,0}$,  $W_\nu^{2,0}$,
$W_\nu^{1,1}$, and $W_\nu^{0,2}$ for $\nu=1,2$  -- they span a base of
the second-rank tensor space.   The tensors $W_\nu^{r,s}$ were divided
by  $W_\nu^{0,0}$ ($i=0,..,2$);  the  tenors with  a twofold  position
weighting  ($W_\nu^{2,0}$)  are  calculated around  the  corresponding
centroid.      All       of      the      other       tensors      are
translation-invariant. Therefore we depict them outside the pattern on
the right-hand-side.  Their eigenvalues are multiplied by  a factor of
$6$ in  order to yield a more  convenient representation. $W^{2,0}_0$:
solid  line;  $W^{2,0}_1$,   $W^{1,1}_1$,  $W^{0,2}_1$:  dashed  line;
$W^{2,0}_2$,          $W^{1,1}_2$,         $W^{0,2}_2$:         dotted
line. \label{fig:beisbart_tensor} }
\end{figure}

In order to give a first example of how the Minkowski valuations
discriminate spatial structures, we consider a few simple
patterns as shown in Figure~\ref{fig:beisbart_examples}. 
In the  case  of a circle  (panel 1) the  curvature centroids
coincide within  the circle's center; the tensors  -- here represented
as  an  ellipse featuring  their  eigenvalues  and the  corresponding
directions  -- are  isotropic.  It  is  useful to  divide the  tensors
$W_\nu^{r,s}$    by     the    corresponding    Minkowski    functionals
$W_\nu^{0,0}$. The  trace of  the normalized tensor  (i.e. the sum  of the
eigenvalues) quantifies to  which extent the morphological information
is  concentrated  around  the  center,  how  large  the  morphological
fluctuations around the  center are. 
\\
The transition to  an ellipse of the same volume  as the circle (panel
2) can be traced either with  the tensors becoming anisotropic or with
the ratio of  the squared surface to the  volume (isoperimetric ratio,
see~\cite{beisbart_schmalzing:webI}).      Furthermore,     as     the
morphological concentration  is decreasing, the traces of  most of the
normalized  tensors  increase,  especially  of $W_2^{2,0}$:  the  more
elongated the ellipse becomes, the higher is its curvature at its very
fringes  along  the horizontal  direction.   The  normalized trace  of
$W_{1}^{1,1}$, on the contrary,  becomes smaller, since the normal and
the local  position vector  cease to be  parallel to each  other.  The
curvature  centroids still  coincide with  each other  because  of the
point  symmetry. --  If we  reduce the  symmetry further  to  a merely
axially symmetric configuration (panel 3), the curvature centroids fan
out along the symmetry axis.  In the case of no symmetry at all (panel
4) they may constitute a triangle.
\\
Finally, suppose  one is given three  convex bodies (or  $d+1$ ones in
$d$ dimensions)  with known Minkowski functionals in  order to combine
them into a  pattern such as indicated in panels 5  and 6.  Given
the  scalar  measures  of  the  whole pattern,  especially  its  Euler
characteristic,  one  can constrain  the  number  of intersections  or
subclumps  in a pattern,  which does  not make  any difference  in our
examples.  For disentangling the positions of the grains, however, one
has  to  know  the  curvature  centroids.   The  centroids  weight  the
subclusters  according  to  their Minkowski  functionals.   Therefore,
whereas from  the volume's point  of view, only the  larger components
contribute significantly,  all subclumps  have the same  weight (Euler
characteristic  $=1$)  for  the  curvature  centroid  $\p_2$.   If  the
subclumps  do not  overlap, one  can indeed  completely  recover their
positions. \\
In Figure~\ref{fig:beisbart_tensor}, we show  all tensors for the last
example in  more detail. Obviously,  there are not  only anisotropies;
rather  different morphological directions  are relevant.   The volume
concentration around the center of  mass is relatively high, since the
volume is  dominated by  the two largest  subclusters, being  close to
each other.  It is  quite the opposite  with $W_{2}^{2,0}$:  $\p_2$ is
relatively far away  from all the clumps, the  curvature around $\p_2$
is  dispersed leading  to a  high value  of tensor's  trace.  The main
relevant direction  connects the pair  of large clumps with  the third
subcluster. The tensor $W_{1}^{1,1}$  is always isotropic, as shown in
Equation~\eqref{eq:beisbart_linear},  its  trace is  connected  to the  volume.
Obviously,  $W_{2}^{1,1}$  and  $W_{1}^{0,2}$ complement  each  other;
indeed  they  add   up  to  a  diagonal  tensor   (recall  the  linear
relationships described in   \cite{beisbart_schneider:tensor}).   Their
eigenvectors' alignment  with the coordinate  axes is due to  the fact
that  all the  subclumps are  aligned parallel  to the  axes  as well.
Otherwise,  $W_{2}^{1,1}$  and $W_{1}^{0,2}$  would  average over  the
orientations of the subclumps.  Finally, $W_2^{0,2}$ is isotropic, its
trace equals  the Euler characteristic.\\ The  detailed description of
patterns given  by the Minkowski  valuations can be used  in practical
applications, as we shall see in the next section.
\section{Physical applications of higher-rank Minkowski valuations}
\label{sec:beisbart_ap}
\subsection{The inner structure of spiral galaxies}
\begin{figure}
\begin{minipage}[t]{.5\linewidth}
\centering
\centering\epsfig{file=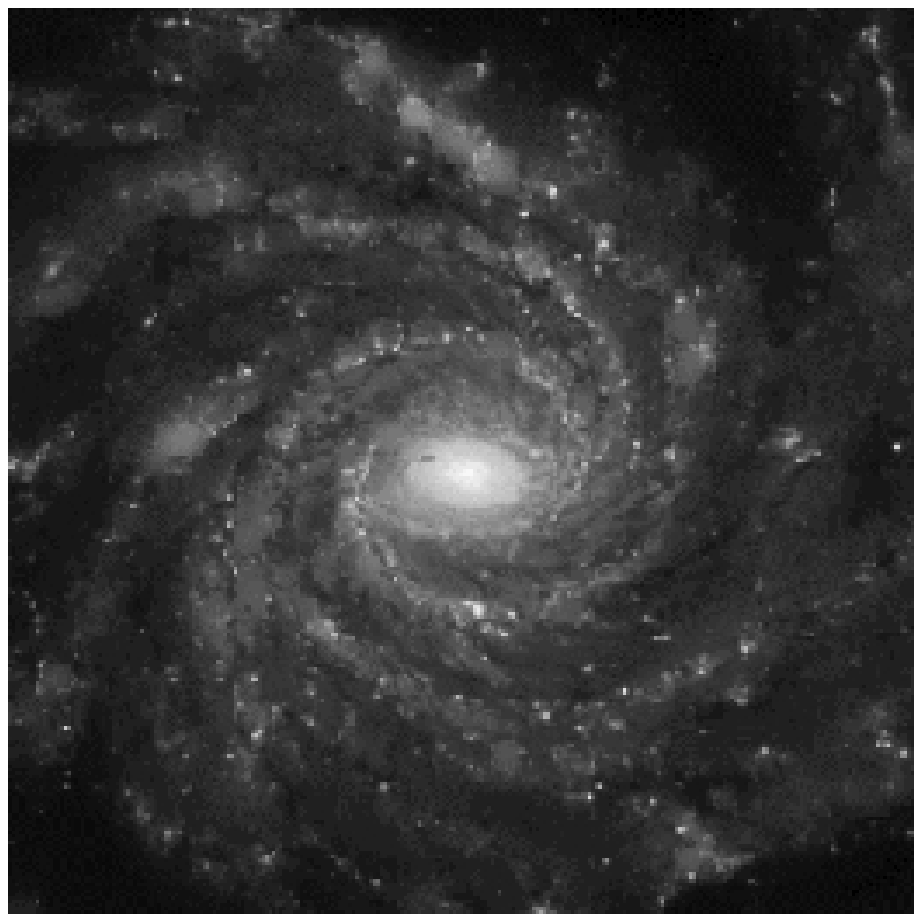,width=6cm}
\end{minipage}\hfill
\begin{minipage}[t]{.5\linewidth}
\centering
\centering\epsfig{file=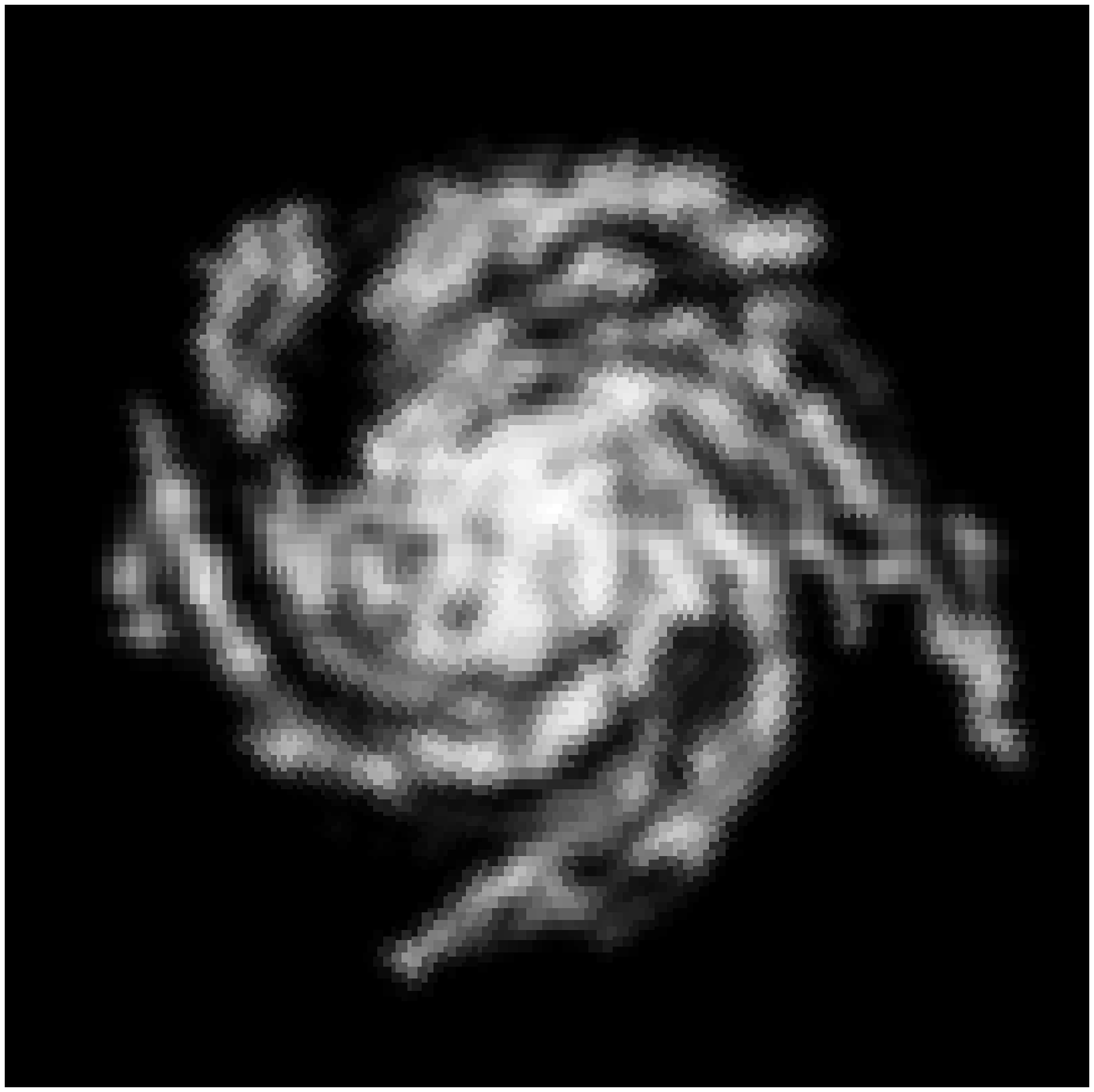,width=6cm}
\end{minipage}\hfill
\centering
\centering
\caption{\label{fig:beisbart_spiral}
The spiral galaxy NGC 1232 (as observed by the VLT/ESO) and a
simulated galaxy following the model by \protect\cite{beisbart_seiden:percolation}.}
\end{figure}

\begin{figure} 
\begin{minipage}[t]{\linewidth}
\centering
\centering\epsfig{file=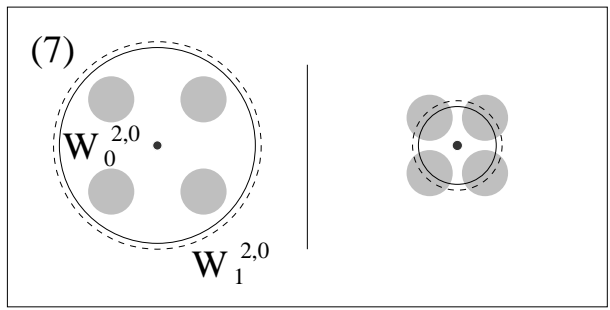, width=8cm}
\end{minipage}\hfill
\centering
\centering
\caption{\label{fig:beisbart_conc}  The tensors  $W^{2,0}_0$  (solid line)  and
$W^{2,0}_1$  (dashed line)  for  a  wide spread  (right  panel) and  a
compact  (left  panel)   configuration  of  circles.  The  eigenvalues
depend quadratically on the separation of the circles.}
\end{figure}

\begin{figure}
\begin{minipage}[t]{.5\linewidth}
\centering
\centering\epsfig{file=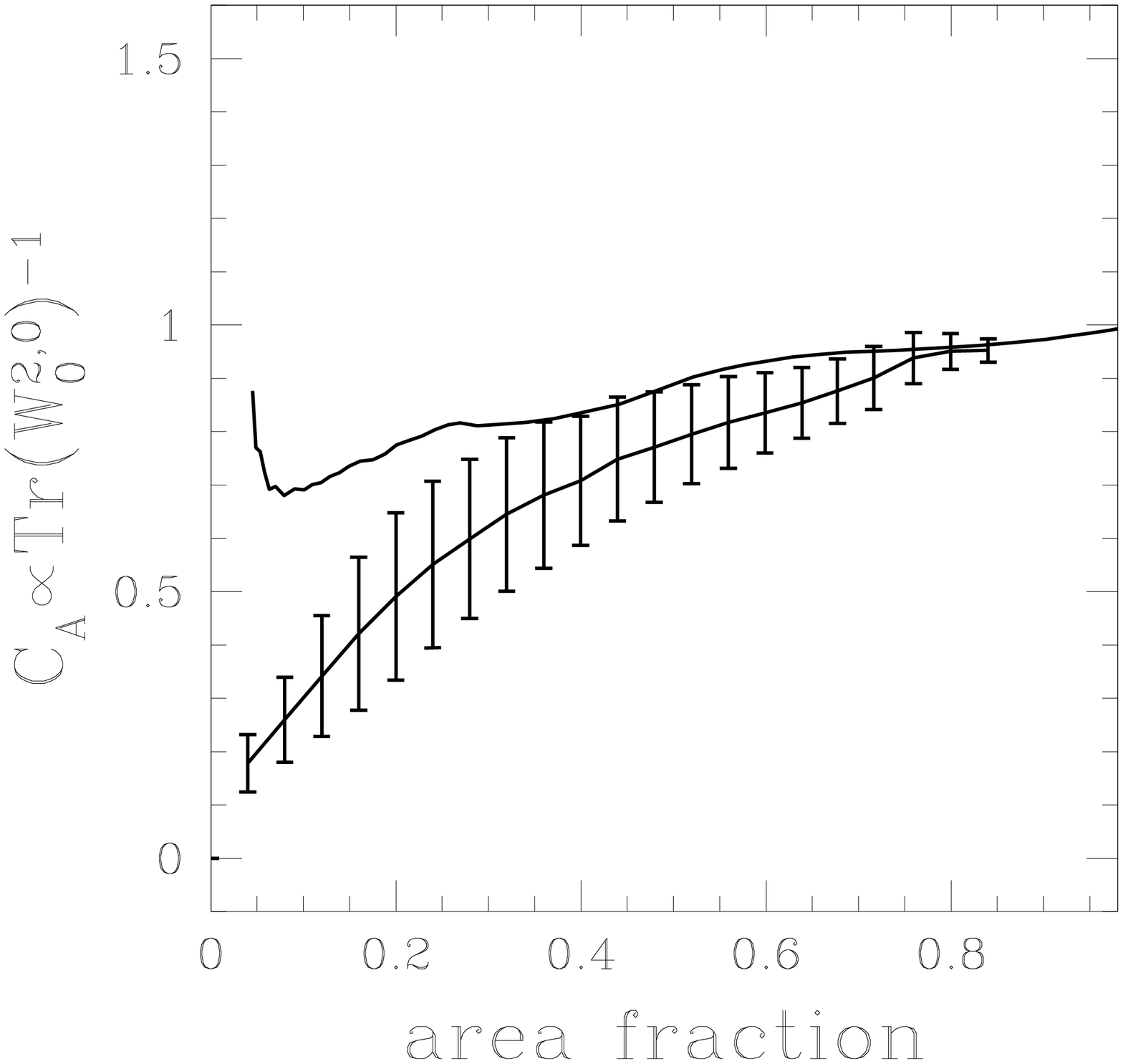, width=6cm}
\end{minipage}\hfill
\begin{minipage}[t]{.5\linewidth}
\centering
\centering\epsfig{file=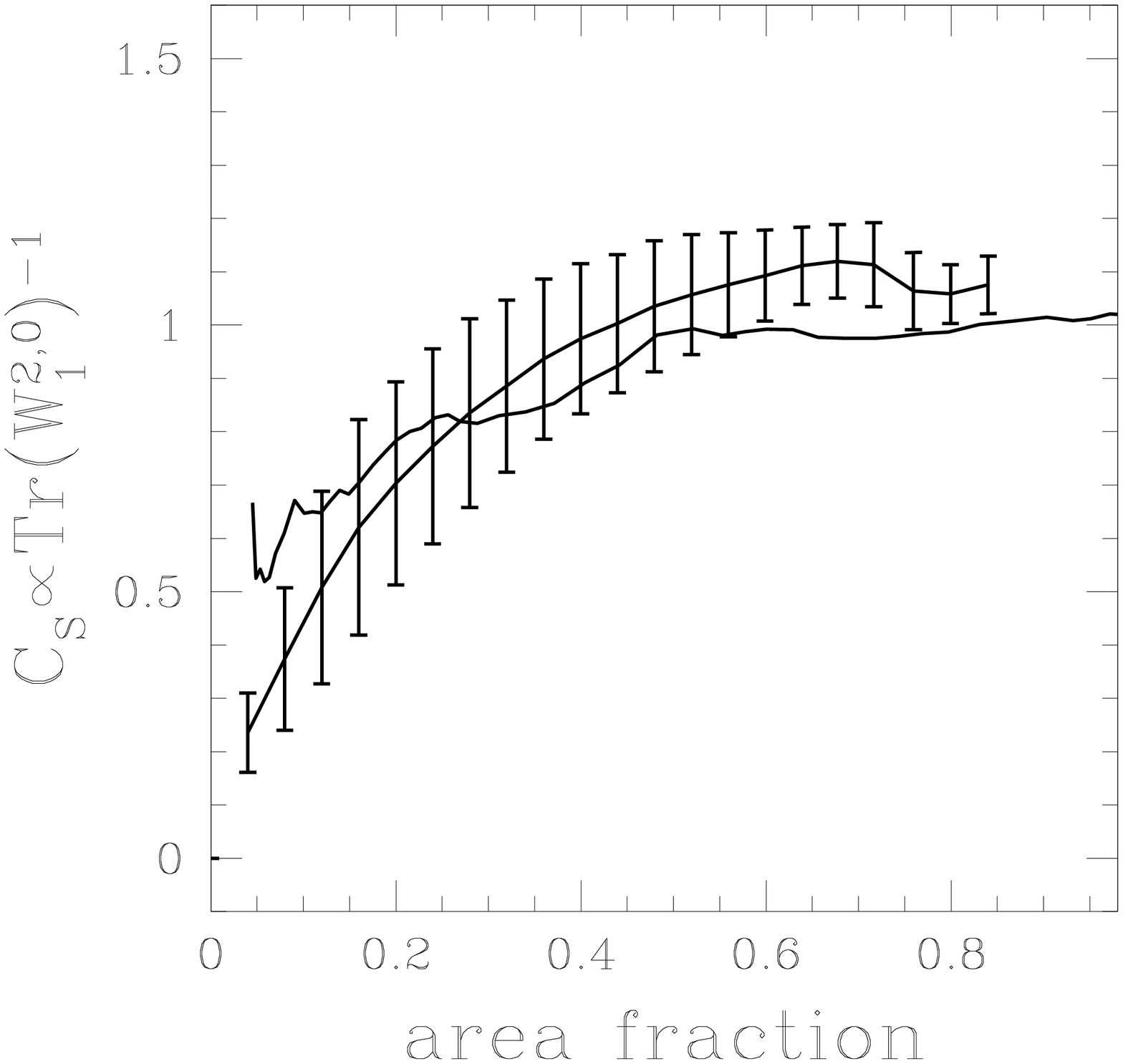, width=6cm}
\end{minipage}\hfill
\centering
\centering
\caption{\label{fig:beisbart_model} Results  of a quantitative  morphometry for
spiral galaxies  using Minkowski tensors.  We consider  the galaxy NGC
1232 (shown in the left panel of Figure~\ref{fig:beisbart_spiral}) and a sample
of model galaxies following a percolation model (an example of a model
prediction     is    displayed     in    the     right     panel    of
Figure~\ref{fig:beisbart_spiral}).   We plot  two concentration  parameters (as
described  in  the main  text)  versus  decreasing surface  brightness
threshold  values,  i.e.\ increasing  surface  area.   The error  bars
indicate $1\sigma$  fluctuations around  the expectation value  of the
model,  whereas the  single line  refers to  NGC 1232.   The parameter
C$_A$ reveals significant differences in the concentration between the
surface brightness profiles of real and model galaxies. The particular
percolation model used thus is incompatible with real data.}
\end{figure}

The  classification of  spiral or  disc galaxies  into  specific types
(the   Hubble  types, e.g.,  {}\cite{beisbart_hubble:classification})   is  an
important  step  towards   their  physical  understanding;  there  are
competing  theoretical pictures  trying  to explain  the formation  of
spiral       arms       \cite{beisbart_lin:spiral,beisbart_gerola:stochastic}.        In
Figure~\ref{fig:beisbart_spiral}  we  display a  real  galaxy observed  face-on
together   with    a   simulation   according   to    a percolation   model   
{}\cite{beisbart_gerola:stochastic}.   So  far  one  has mainly  relied  on  an
assignment by eye  in order to classify observed  disc galaxies and to
compare     observational     data     with     theoretical     models
\cite{beisbart_naim:comparative}.  One  of the reasons for this  is that spiral
arms are  easily detected  by the  human eye, but  are rather  hard to
quantify.  Here we  use Minkowski valuations to reveal  aspects of the
inner       structure        of       spiral       galaxies       (see
\cite{beisbart_russell:spiral,beisbart_block:spiral,beisbart_kornreich:asymmetry}
for some discussion of how to quantify morphometric features of spiral
galaxies, and especially \cite{beisbart_abraham:automated,beisbart_abraham:asymmetry}).
\\
In order to  render an optical galaxy image  accessible to a Minkowski
analysis,  it  is helpful  to  understand  it  as a  two-dimensional
surface  brightness  field  $u(\x   )$.   We  may then estimate  the  Minkowski
valuations of the excursion  sets $M_v=\{\x |u(\x )>v\}$.  Contrary to
large scale  cosmological data (e.g.\ the distribution  of galaxies or
clusters of galaxies), such a density field is neither homogeneous nor
isotropic;  rather it reveals  a point  of considerable  symmetry: the
center of mass.  We expect  the curvature centroids to coincide within
this point and therefore concentrate on second-rank Minkowski tensors.
\\
In  order to  detect  a  possible overall  anisotropy,  we divide  the
smaller  eigenvalue of  $W^{2,0}_0$  by the  larger  one; this yields  the
parameter $X$.   Since $X\leq 1$ by definition,  $\epsilon \equiv 1-X$
can  serve as a  measure of  the anisotropy.   Such an  anisotropy may
reflect a  physical anisotropy, induced,  for instance, by the presence  of two
massive  spiral arms, but  may also  indicate that  the galaxy  is not
observed face on, but rather under an inclination angle.  
\\
As  the tensors $W^{2,0}_\nu$ ($\nu=0,..,2$) weight the volume or the
surface, respectively, 
with $\x\x$, their eigenvalues  are increased, if the configuration is
more widespread  (see Fig.~\ref{fig:beisbart_conc}). This fact can  be used to
quantify the  concentration of  a density field  around its  center of
mass, by calculating the trace of $W^{2,0}_\nu$.
Therefore another characteristic feature  of disc galaxies, namely the
way  in which  the  radial surface  brightness distribution  declines
(with increasing radius), can  be analyzed by scanning through various
threshold values $v$.  We applied (suitably normalized) concentration
parameters $P_\nu$  (built from $\trace  (W^{2,0}_\nu)$) to the  images of
both real galaxies (C.  Moellenhoff, private communication) as well as
to galaxies       produced       by       a       percolation       model
(\cite{beisbart_seiden:percolation};   see   {}\cite{beisbart_dahlke:diploma}  for
details).   These morphological  descriptors reveal  strong luminosity
concentration for real galaxies, which  could not be reproduced by any
of  the  model  galaxies  (see  Fig.\ref{fig:beisbart_model})  with  reasonable
choices for the model parameters.
Using the  morphology of observed disc galaxies,
one  can therefore rule  out the  percolation model  (at least  in its
simple form)  as a viable attempt  to explain the  formation of spiral
arms. 
\subsection{The morphology of galaxy clusters}
As mentioned in Section~\ref{sec:beisbart_mot},  the morphology of clusters may
constrain  the values of  the cosmological  parameters. Intuitively,  one may expect clusters  in
cosmological  low-density models  to have  less substructure  than in
high-density  models~\cite{beisbart_richstone:cluster}.  Using Minkowski  valuations, we  check whether
this   effect    can   be   observed   for    the   cosmological   GIF
simulations~\cite{beisbart_bartelmann:arcIV}.    We   analyze  the   clusters'
projected     mass     distributions      such     as     shown     in
Figure~\ref{fig:beisbart_problem}.  Observations from the gravitational lensing
through       clusters~\cite{beisbart_schneider:book,beisbart_fischer:lens}       trace
approximately the same  cluster information.  \\ In order  to make the
simulation data, which consist of ``particles tracing matter'', accessible
to the  Minkowski valuations, we  construct a pixelized  density field
$u(\x)$ and smooth it  with a Gaussian kernel of width\footnote{Length
scales are  given in  units of $\mpch$.   One Megaparsec  (Mpc) equals
about 3.26 million light years,  the constant $h\sim 0.65$ accounts of
the  uncertainty in  the  measurements of  the  Hubble constant.} 
$\lambda=0.3\mpch$.   The  resulting  field $u_\lambda(\x)$  contains  the
overall  shape  of  the  cluster.   Again we  estimate  the  Minkowski
valuations of  the excursion  sets, which contain  all pixels  above a
given  density threshold  $v$. Our  method to  estimate  the Minkowski
functionals    from    the    grid    is    based    upon    Crofton's
formulae~\cite{beisbart_serra:morphology,beisbart_schmalzing:beyond,beisbart_beisbart:diss}.
For each cluster we average a function of the Minkowski valuations $f$
over   different   density   thresholds:  
\begin{equation}
\frac{1}{\left(   u_\lambda
\right)_2-\left(    u_\lambda   \right)_1}\langle    f\rangle   \equiv
\int_{\left(  u_\lambda  \right)_1}^{\left(  u_\lambda \right)_2}  \dd
u_\lambda  f
\end{equation}
 in order  to obtain  a compact  set of  parameters. Here
$\left( u_\lambda  \right)_1$ and $\left(  u_\lambda \right)_2$ denote
appropriately  chosen integration  limits. We  consider five  types of
substructure.
\begin{enumerate}
\item        The        {\em       clumpiness}        C $\equiv\sqrt{\langle
\left(\frac{1}{\pi}W_2-1\right)^2\rangle}$   counts    the   number   of
subclumps   in   quantifying  how   much   the  Euler   characteristic
$W_2/ \pi$ deviates from one.
\item The  construction of the {\em asymmetry}  parameter A is  based on the
fact that curvature centroids which  do not coincide with each other,
indicate  asymmetry.  Therefore,  we  calculate the  perimeter of  the
triangle    constituted    by     the    curvature    centroids,    $2
W_1\left(\Delta\left(\p_i\right)\right)$,      A$\equiv\langle
W_1\left(\Delta\left(\p_i\right)\right)\rangle$.
\item The ratio of the tensors' eigenvalues, $\gamma_1 \left( W_\nu^{r,s}\right)/\gamma_2\left( W_\nu^{r,s}\right)$ tells
us  to  which  extent   {\em anisotropies}  are  present. Here we
consider the tensor $W_1^{2,0}$ and define  X$\equiv  \Big\langle
\gamma_1 \left(W_1^{2,0}\right)/\gamma_2\left(W_1^{2,0}\right)\Big\rangle$.
\item In general the curvature  centroids $\p_\nu$ wander in space if
the density threshold is  varied; their fluctuations therefore measure
the  {\em shift of  morphological properties}.  Focusing on  $\p_1$ we
define        S$\equiv        \langle       \left(        \p_1-\langle
\p_1\rangle\right)^2\rangle$.  This   parameter  is  related   to  the
centroid   variation    used   in   cluster    investigations   before{}\cite{beisbart_mohr:x-ray-cluster}.
\item 
Likewise, the  eigendirections (parametrized by an  angle $\alpha$) of
the  tensors   are  changing;  this   generates  the  {\em   twist  of
morphological}  properties.  Restricting  ourselves to  $W_1^{2,0}$ we
employ      the      parameter     T$\equiv\langle\left(\alpha-\langle
\alpha\rangle\right)^2\rangle$.
\end{enumerate}
 For more technical details, see~\cite{beisbart_beisbart:diss}.
\begin{figure} \centering
\begin{minipage}[t]{.33\linewidth}
\centering
\centering\epsfig{file=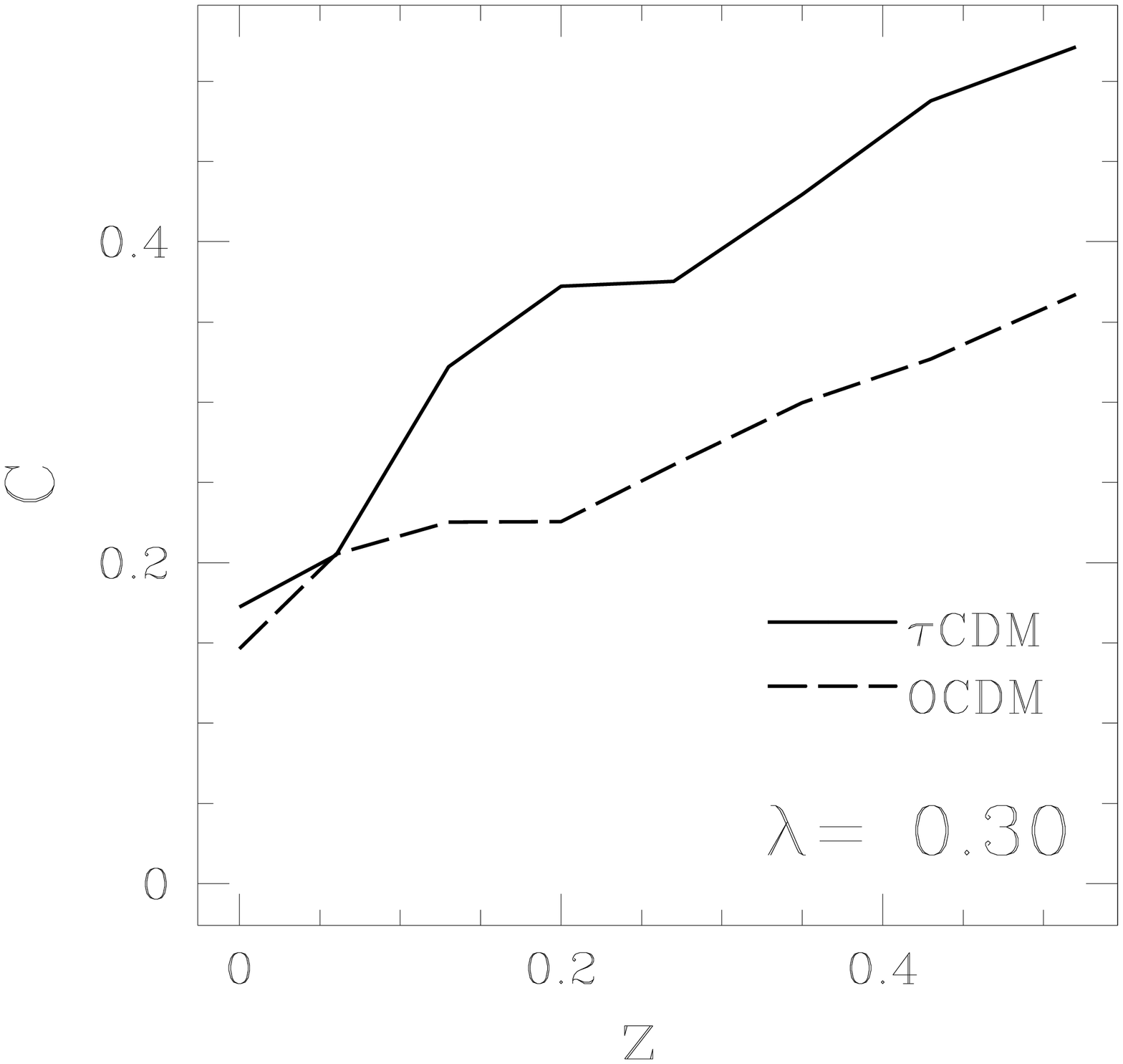,width=4cm}
\end{minipage}\hfill
\begin{minipage}[t]{.33\linewidth}
\centering
\centering\epsfig{file=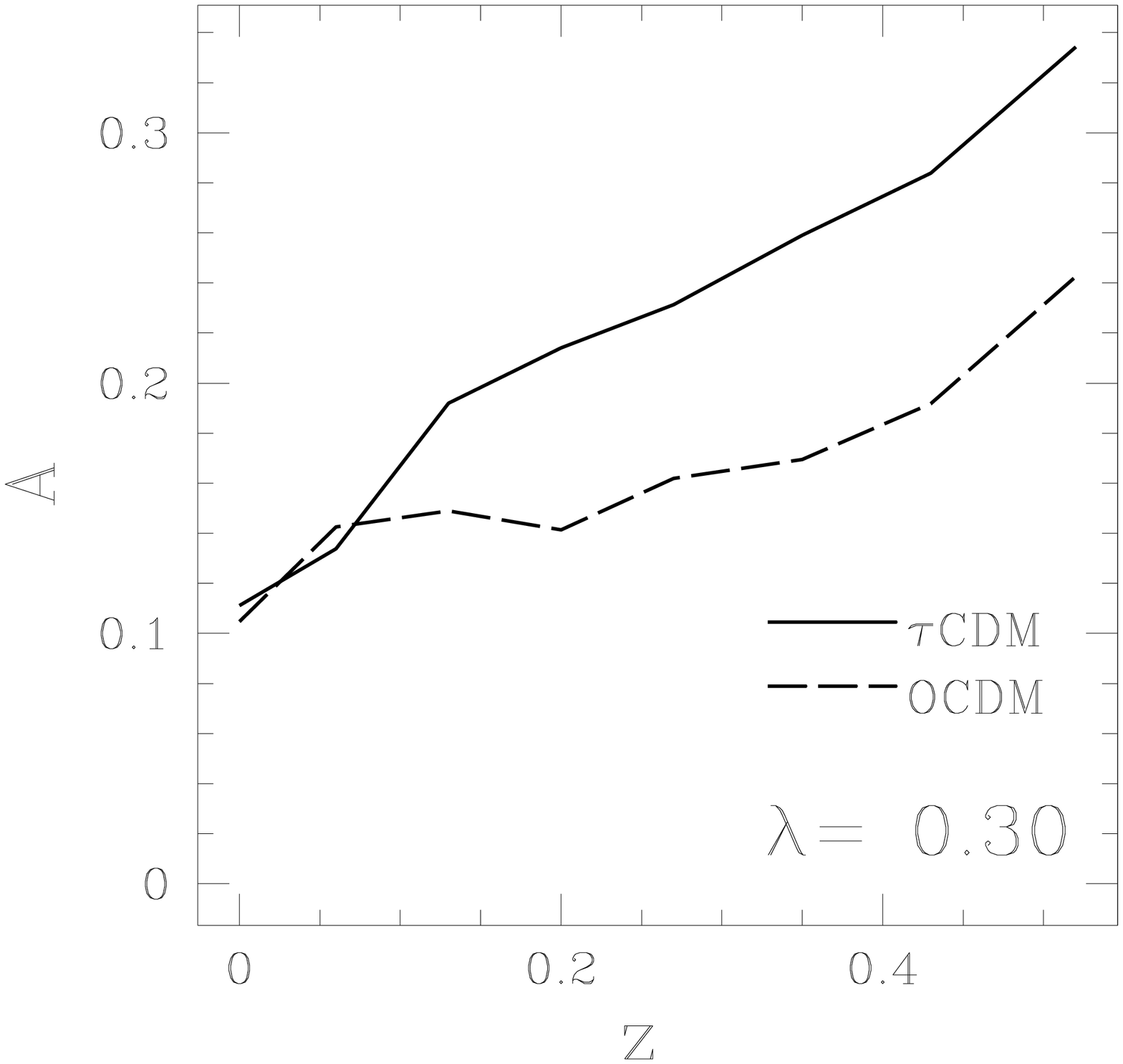,width=4cm}
\end{minipage}\hfill
\begin{minipage}[t]{.33\linewidth}
\centering
\centering\epsfig{file=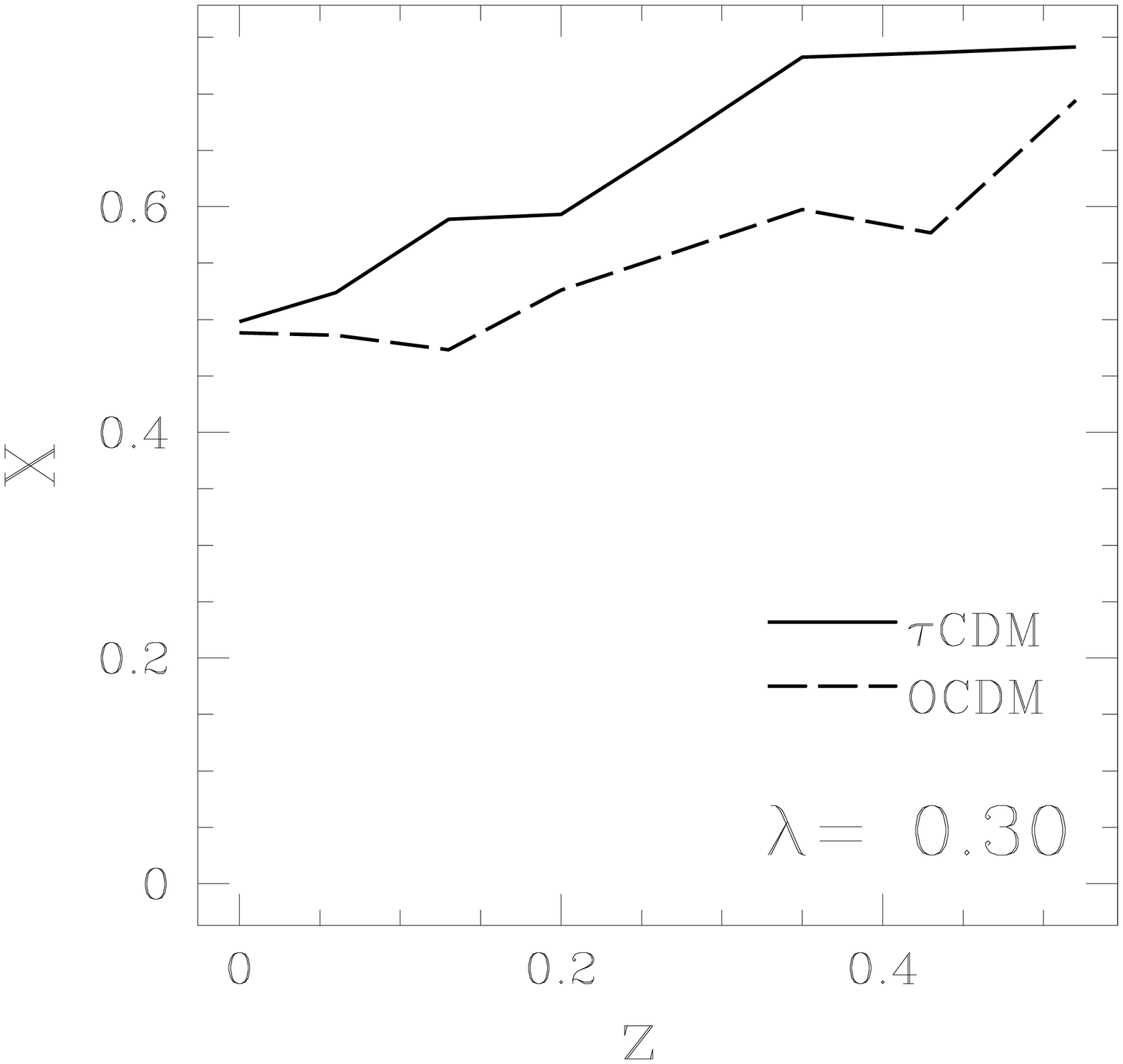,width=4cm}
\end{minipage}
\begin{minipage}[t]{.15\linewidth}
\end{minipage}
\begin{minipage}[t]{.33\linewidth}
\centering\epsfig{file=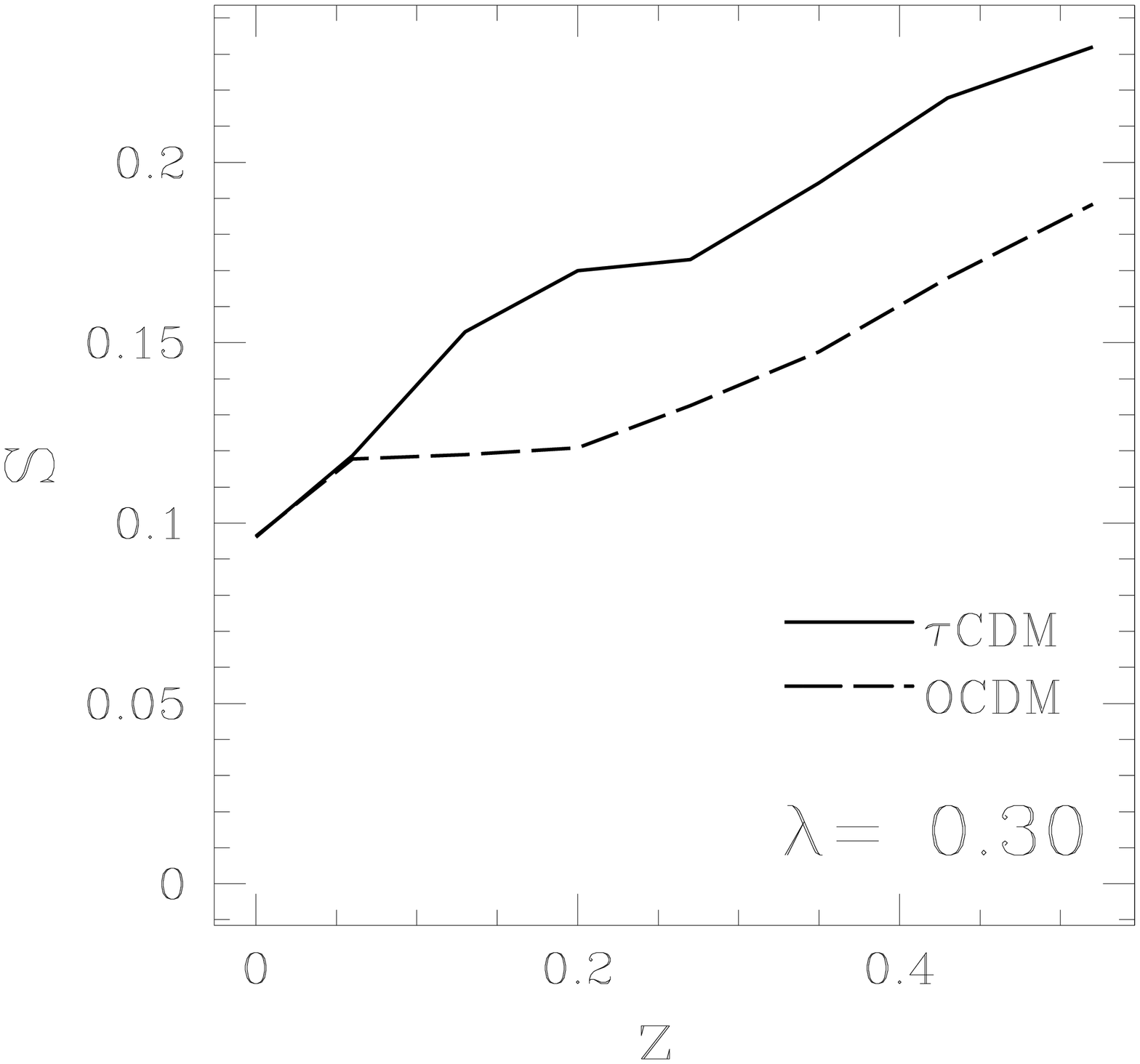,width=4cm}
\end{minipage}
\begin{minipage}[t]{.0\linewidth}
\end{minipage}
\begin{minipage}[t]{.33\linewidth}
\centering\epsfig{file=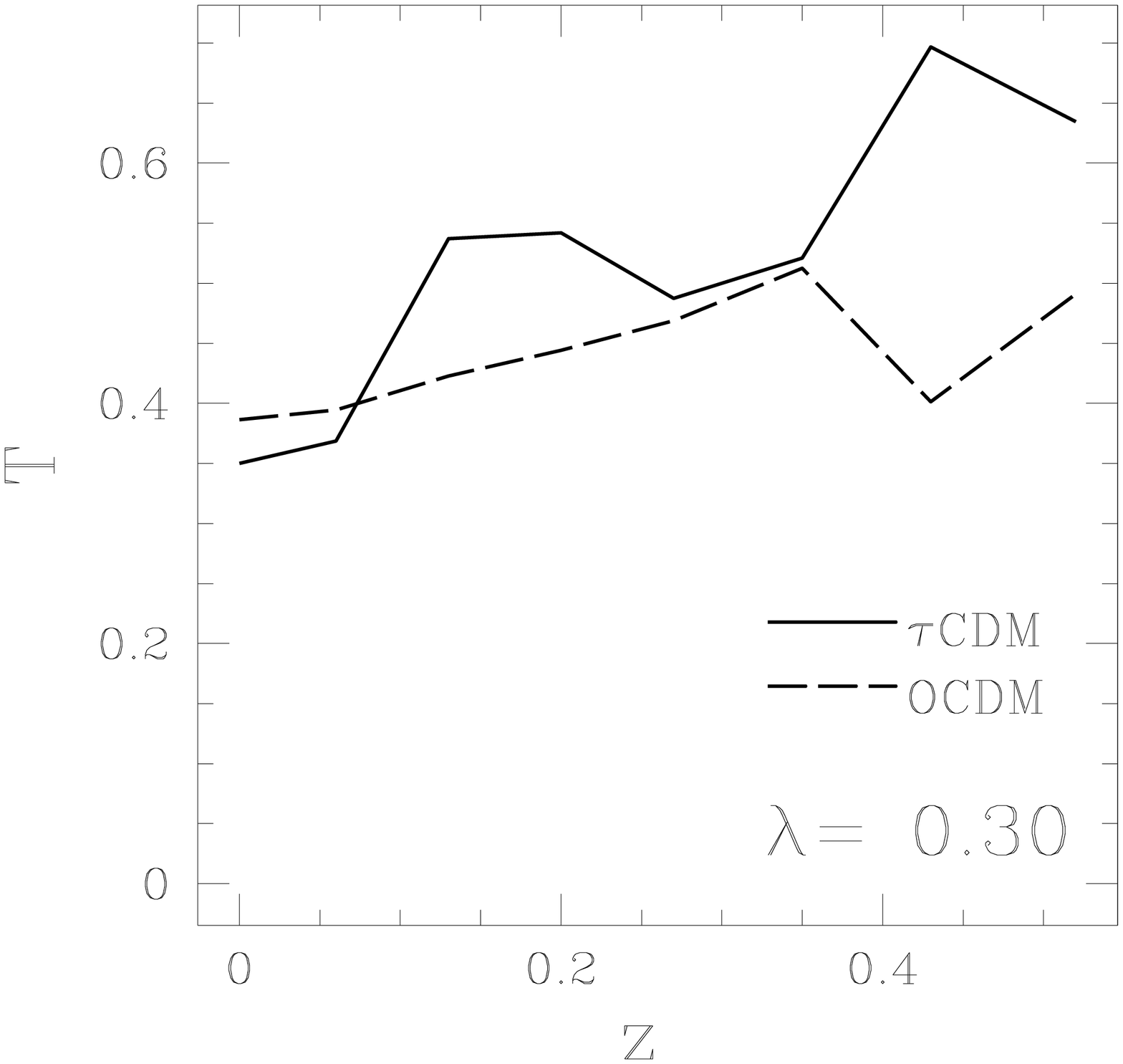,width=4cm}
\end{minipage}
\begin{minipage}[t]{.15\linewidth}
\end{minipage}
\centering   
\centering   
\caption{ The average morphological  evolution of galaxy clusters from
the GIF  simulations. Our morphological order  parameters clumpiness C,
asymmetry  A,  anisotropy  X,  shift  of morphology  S  and  twist  of
morphology T are shown versus time.  The time is given in terms of the
observable  cosmological redshift.  Redshift  $z=0$ means  the present
day, higher  redshifts lead back into  past.  In order  to get typical
results for  two cosmological models (low-density model  with an open
geometry,   called  open   Cold  Dark   Matter,  OCDM:   dashed  line;
high-density model  with a massive $\tau$  neutrino, named $\tau$CDM:
solid line)  we average  over about $15$  clusters and  $3$ orthogonal
projections per cluster  within each model.  As the  results show, our
morphological order parameters  quantitatively feature the differences
between  the  models which  are  in  accordance  with the  theoretical
predictions. For details on the cosmological Cold Dark Matter models
see {}\protect\cite{beisbart_peacock:cosmological}\label{fig:beisbart_gif_res}}
\end{figure}
For  each  time period (or  cosmological  redshift) we  estimate  the  order
parameters  for three  projections per  cluster and  average  over all
clusters  in  one  cosmological  model  in  order  to extract  a  typical
morphology.  In Figure~\ref{fig:beisbart_gif_res},  we display the evolution of
the  averaged morphological parameters  vs. redshift.   One recognizes
immediately a  relaxation process, leading to  less substructure (and
lower values of the morphological order  parameters) as time goes by  (from right to
left).   Furthermore,  clear   differences  between  the  cosmological
background models arise.  Indeed the theoretical predictions according
to which  the low-density  clusters (OCDM) exhibit  less substructure,
are  confirmed for  the past;  for redshift  $z=0$,  the differences
disappear.\\ All of our order parameters show a similar behavior. The
reason probably is that we  average over different types of relaxation
processes.   An   investigation  of  single   clusters  should  reveal
differences between, e.g., a bimodal merger dominated by two colliding
subclumps and a more isotropic mass infall.
\subsection{The geometry of the electric charge distribution in molecules}
\begin{figure} \centering
\centering\epsfig{file=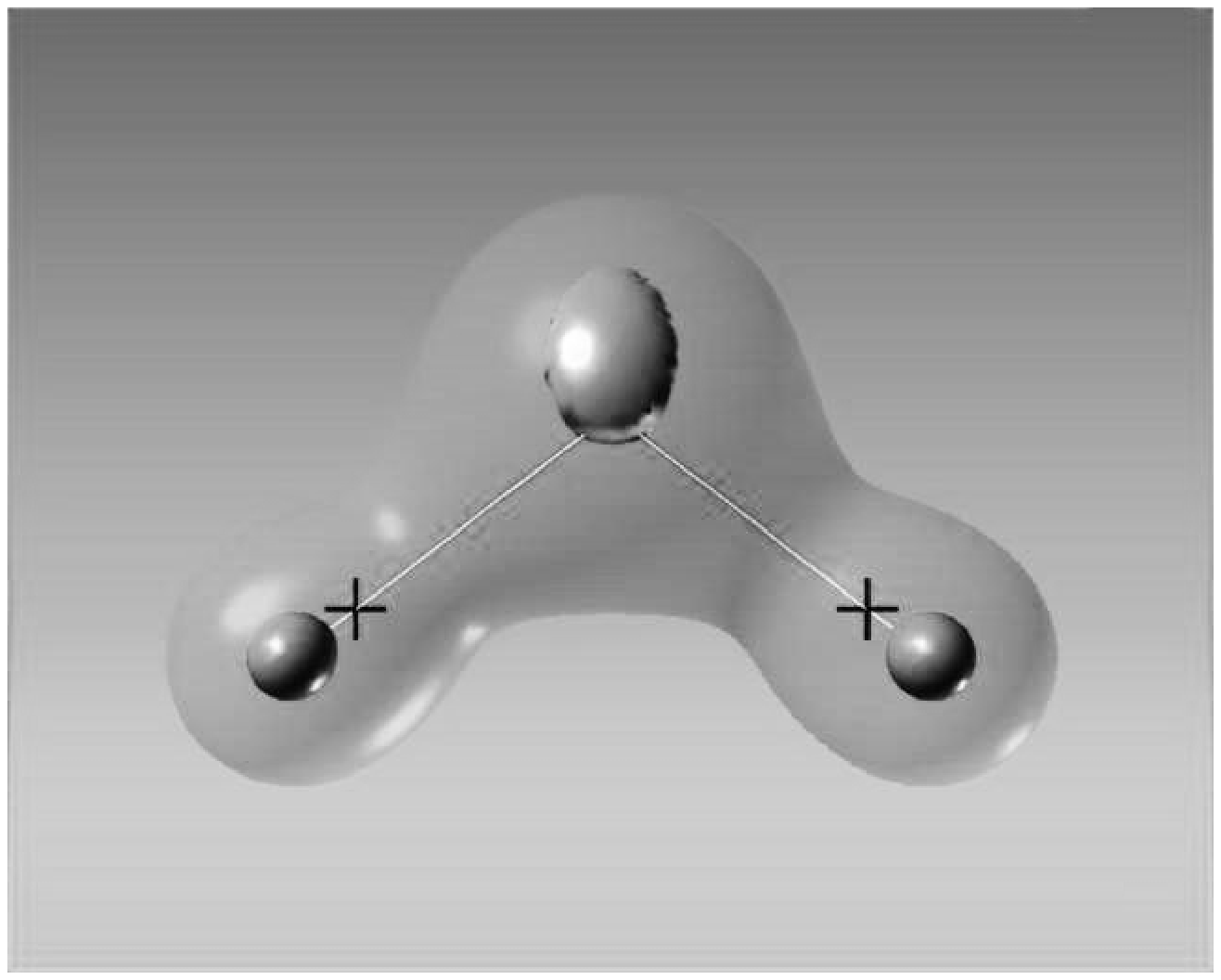,width=6cm}
\centering\epsfig{file=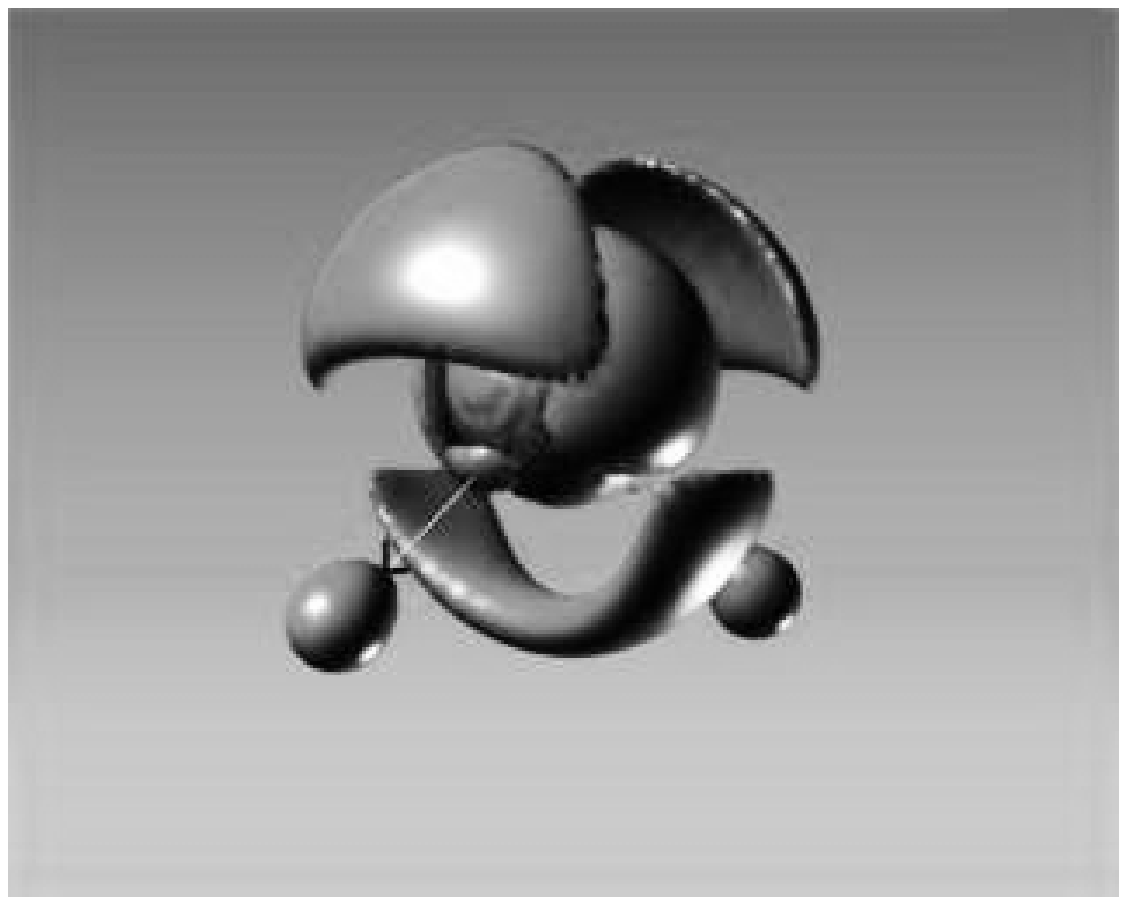,width=6cm}
\caption{\label{fig:beisbart_atoms}The  water   molecule  H$_2$O.  The
bright lines denote  the bonds connecting the H-atom  with both of the
O-atoms.  The crosses  mark the  points where  the  zero-flux surfaces
separating  the  atoms  intersect  with  the  bonds.   Moreover,  some
iso-surfaces  of  the  Laplacian  of the  charge  density,  $\triangle
\varrho$,      are      shown.       The     images      are      from
http://www.nas.nasa.gov/$\sim$creon/papers/mgms96/            (courtesy
C.~Levit).   The  panels  show  different  projections  and  focus  on
different iso-surfaces.}
\end{figure}

An important implication of the atom hypothesis is that the structures
of  mole\-cules are  to be  explained  in terms  of their  constituent
parts, the  atoms.  Heuristic principles based  upon this assumption
work quite  well in chemistry. However,  if one tries  to justify this
hypothesis in  the framework  of quantum mechanics,  the notion  of an
individual  atom  seems  to   disappear,  since  the  negative  charge
distributions of the atoms  merge, blurring the electron clouds which
define the  single atoms.  
\\  
Bader \cite{beisbart_bader:quantum}
proposed an elegant method to
recover the  notion of  atoms in molecules.   Since, from  the quantum
mechanical point of view, the charge density profile $\varrho(\rb)$ is
the only relevant quantity, its geometry and topology must contain the
notion  of  atoms.   Using  the critical  points  of the  density
distribution  $\varrho$, where $\nabla  \varrho =  0$, one  can define
atoms  as local  density  maxima together  with  a surrounding  domain
enclosed  by zero-flux  surfaces.  This  geometrical approach  can be
extended to chemical bonds and has been justified on physical grounds.
Moreover, it provides us with means to visualize molecules in a proper
way  using  the charge density  profile,  see  Figure~\ref{fig:beisbart_atoms} for  an
illustration.  A distinguished role is played by the surface $\nabla^2
\varrho=0$ which separates the zone where chemical reactions are likely to
happen from the rest of space.
\\
Obviously,  this is  an area  of research  where physics  and geometry
strongly  interact,  and  the  applications of  higher-rank  Minkowski
valuations  may  yield useful  information  on  the  structure of  the
chemical bonding.  First, the morphology of the whole molecules can be
simply described in  terms of the Minkowski valuations  of the density
field.  The atoms,  which Bader defines \cite{beisbart_bader:quantum},
may  not be compact  in every  case, so  the single  atoms can  not be
understood in  terms of the  Minkowski valuations.  However,  the fact
that the atoms' positions are specified in Bader's theory gives us the
chance  of probing  the  molecular structure  with  the Boolean  grain
method where the atoms'  positions are decorated using spheres (either
monodisperse or with radii mirroring physical quantities).  Note, that
at  this  point one  really  needs tensors,  since  in  this case  the
locations  and   the  orientations   of  subsystems  are   of  crucial
importance; and  this local information  lies beyond the scope  of the
scalar  measures.  Bader's theory  constructs also  a graph  of bonds,
that may be analyzed with Minkowski valuations straightfowardly.
\\
But given the fact that  both the molecules and the constituting atoms
are defined in terms of geometry,  one may go one step further and try
to discover  phenomenological principles  allowing one to  predict the
shape of molecules given their constituents.
\\
These   endeavors   clearly   are    beyond   the   scope   of   this
article. However, in order to illustrate to some extent how useful the
Minkowski valuations may prove in  this area, we consider briefly some
of the Minkowski tensors for the H$_2^{+}$ ion.
\begin{figure} \centering
\centering\epsfig{file=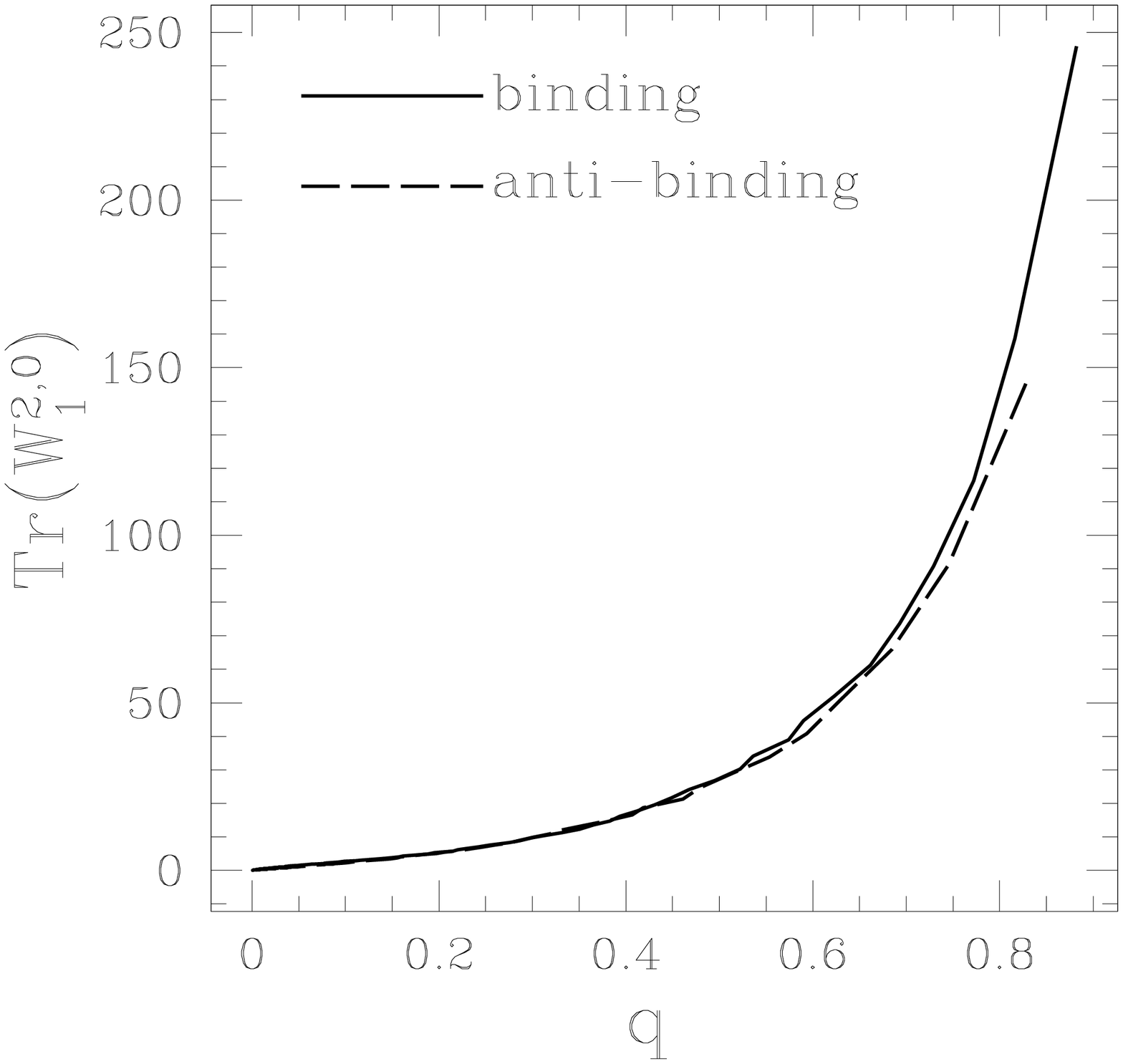,width=6cm}
\centering\epsfig{file=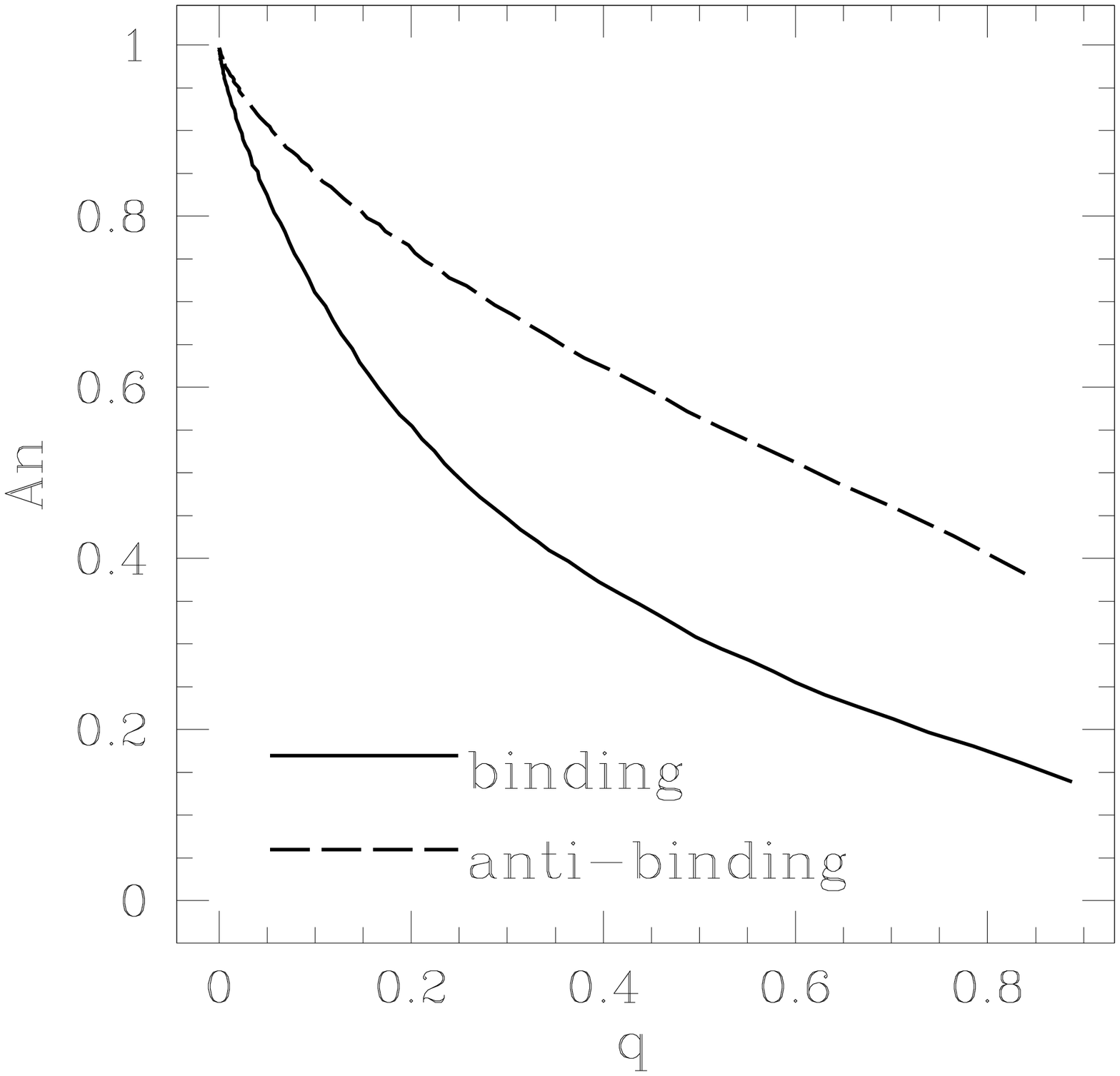,width=6cm}
\caption{The morphology of the bonding and the anti-bonding states of
the  electron   within  the  $H^+_2$  ion.   We   explore  the  charge
distribution of  the electron via  the Minkowski tensors  of excursion
sets from the  density field. In the left panel, we  show the trace of
$W^{2,0}_1$ as a  function of the charge fraction q enclosed within the
density contour.   In the right  panel, we consider the  anisotropy of
the     charge     distribution.      It     is     quantified     via
$An\equiv\left(\left(W^{2,0}_1\right)_{xx}-\left(W^{2,0}_1
\right)_{yy}\right)/\trace\left(W^{2,0}_1\right)$.   The anti-bonding
state,  shown as the  dashed line  is much  more anisotropic  than the
bonding  state   which  wraps  both  atoms  with   a  smooth  electron
hull.\label{fig:beisbart_bonding}}
\end{figure}
\\
In order to  treat the H$_2^{+}$ ion consisting of  two protons and an
electron in  the framework of  quantum mechanics, one  often considers
the  protons  as fixed in  space (Born-Oppenheimer
approximation)   thereby  reducing   the  problem   to   an  effective
one-particle problem  of an electron  moving in the potential  of two
positively   charged   nuclei.     A   reasonable   ansatz   for   the
quantal wave function of the electron -- at least suitable
to illustrate  what is  basically happening --  is a  superposition of
quantum states where  the electron is bound to  either of the protons.
To be more precise, let us  consider the protons as being separated by
their equilibrium distance of approximately two Bohr radii\footnote{In
the  following  we  consider   spatial  distances  in  units  of  Bohr
radii. The Bohr  radius $a_0$ marks the typical  scale of the hydrogen
ground  state wave function.   For an  introduction into  the physical
treatment  of the  H$_2^{+}$  ion one  may  consult, e.g.   Complement
G$_{XI}$ in \cite{beisbart_cohen:qmii}.}. Let us label the protons with $1$ and
$2$ and  let $\psi_i$ denote the  unperturbed (well-known)
hydrogen ground  state wave  function of an  electron bound  to proton
$i$ for   $i=1,2$. For symmetry reasons, only the linear combinations
\begin{equation}
\psi_{a}  =\frac{1}{\sqrt{2}}\left(\psi_1  -  \psi_2\right)\quad  {\rm
and}\quad \psi_{b} = \frac{1}{\sqrt{2}}\left(\psi_1 + \psi_2\right)
\end{equation}
are of interest. Since $\psi_b$ is favored on energetic grounds, it is a
bonding state; consequently  $\psi_a$ is called antibonding state.
\\
The  morphology  of  the  electron charge  distribution  is expected
to distinguish
between  both states.   Let us consider excursion  sets of  the
three-dimensional  electron  charge  density  profile  $\varrho=\vert
\psi\vert^2$      for     both      states.      In
Figure~\ref{fig:beisbart_bonding}  we  display the surface tensor  $W_{1}^{2,0}$ as  a
function of the charge fraction  enclosed within the excursion set. We
start  from the  very peaks  of the  charge distribution  (high charge
density  thresholds) and move  on to  lower charge  density thresholds
enclosing a  significant part  of the whole  electron charge.   The
trace of the tensor $W_1^{2,0}$  shows that the iso-surfaces of the bonding
state  are  slightly more  extended  than  those  of the  anti-bonding
state. But  on the  whole, the differences  are small: the
concentration of  the charge  distribution is basically  determined by
the distance of  the protons.  If we  ask for morphologically
relevant directions, the differences between both states are much more
perspicuous.  The anisotropy of the azimuthal symmetric quantum states
can                  be                 quantified                 via
\begin{equation}
An\equiv\left(\left(W_{1}^{2,0}\right)_{xx}-\left(W_{1}^{2,0}\right)_{yy}\right)/
\trace    \left(W_{1}^{2,0}\right)\;\;\;,     
\end{equation}
the    protons    lying    at
$\left(x,y,z\right)=\left(-1,0,0\right)$   and   $\left(1,0,0\right)$,
respectively.  For  small charge fractions, the  tensor $W_1^{2,0}$ is
dominated by the peaks  around the positively charged protons yielding
a  maximum  anisotropy,  but  no  essential  difference  between  both
states. However,  as we  move to lower  density thresholds  and higher
charge fractions, the bonding state isotropises much more quickly
than the anti-bonding one.
In the bonding state, therefore,  the electron forms a cloud spreading
out  over  both  atoms  in  a  quite isotropic  way,  whereas  in  the
anti-bonding state, the signature  of the two distinct protons remains
still  visible.   The  chemical  bond  therefore is  mirrored  by  the
geometry  of  the electron  density  distribution,  as quantified  via
Minkowski tensors.
\\
This example is but an  elementary first illustration of the nature of
the chemical bond.  In the spirit of Bader's  theory, one can possibly
go   much   further   in   order   to   consider   single   atoms   in
molecules. Additivity  can serve as  a guiding principle,  since Bader
found that there are  stable atom configurations which persist through
different  types  of molecules,  presenting  a  sort  of rigid  module
movable as a whole; this fact is well-known from chemistry, of course.
\section{Density functional theory based on Minkowski valuations}
\label{sec:beisbart_an}
\begin{figure}
\begin{minipage}[t]{.99\linewidth}
\centering
\epsfig{file=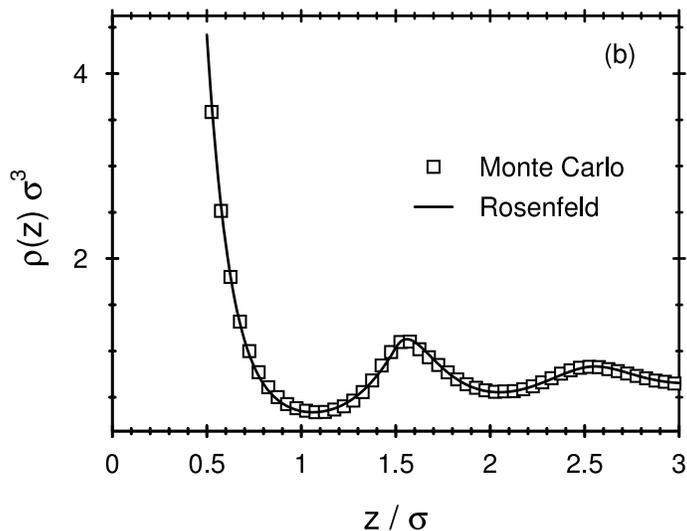,width=9cm}
\end{minipage}\hfill
\caption{ The density  profile of a hard sphere fluid  close to a hard
wall  for a bulk  density $\rho  (2\sigma)^3 =  0.715$ calculated  with the
Rosenfeld functional  (solid line)  as compared to  simulation results
(squares,       Figure       from       R.~Roth,       1999,       see
\cite{beisbart_roth:depletion}). $\sigma$ denotes  the diameter of the
spheres.
\label{fig:beisbart_ro}}
\end{figure}

Having shown the morphometric versatility of MVs, we conclude our
discussion of applications by considering systems where integral
geometric quantities enter a physical approximation. The quantities
that come into play at this point generalize the Minkowski valuations
further and may indicate a new direction of interest.
\\
Density  functional  theory  became   the  standard  method  to  study
inhomogeneous classical fluids in the  last 20 years. The main feature
of such physical  systems is the spatial variation  of the equilibrium
density  $\rho_{eq}(\x)$  which minimizes  the  free energy  $\Omega$.
Density functional theory is  based on the exact result  that the free energy
of  the   inhomogeneous  fluid  can  be  expressed   as  a  functional
$\Omega[\rho(\x)]$  of  the density  $\rho(\x)$.   Thus, all  relevant
thermodynamic properties such as surface tensions or phase transitions
of  confined fluids  can be  calculated  from $\Omega[\rho_{eq}(\x)]$;
even equilibrium  correlation functions that  describe the microscopic
structure of inhomogeneous fluids  can be determined by derivatives of
the functional.
\\
In  general, the  form  of the  functional  $\Omega[\rho(\x)]$ is  not
known; deriving  the exact functional would be equivalent
to solving  the statistical mechanics for fluid  systems exactly which
seems  to be an impossible task. Therefore,  one has  to restrict  oneself to
reasonable  approximations for $\Omega[\rho(\x)]$.   For convenience,
one   separates   an   excess   intrinsic   free   energy   functional
${\f}_{ex}[\{\rho_i\}]$ from exactly known external contributions
\begin{equation}
\label{Omega}
\Omega[\rho(\x)]=\int       \dd       \x       \;       \left(
{\f}_{ex}[\rho(\x)]   \;  +\;   k_BT  \rho(\x)   \left(  \ln
\Lambda^3  \rho(\x)  -1\right)   \;  +  \;  \rho(\x)  \left(
V(\x) - \mu \right) \right)
\end{equation}
where $V(\x)$  is an external potential, $\mu$  the chemical potential,
$\beta=1/(k_BT)$  the  inverse  temperature  $T$,  and  $\Lambda$  the
thermal
wavelength of the fluid. $k_B$ denotes the Boltzmann constant.
\\
The most elaborated and  reliable functional for hard-sphere fluids is
that  developed  by  Rosenfeld  {}\cite{beisbart_rosenfeld:free} based  on  the
local scalar and vector-valued Minkowski functionals
\begin{equation}\label{eq:minkfunc}
W_\nu^{p,q}[\rho(\x)]    \equiv    \frac{1}{\nu   \binom{d}{\nu}}
\int_{\partial          K}          \dd          S^{d-1}s_{\nu-1}
\left(\kappa_1,..,\kappa_{d-1}\right) \x^p \n^q \;\; \rho(\x) \;.
\end{equation}  
In  contrast  to the  definition  \eqref{eq:beisbart_alesker} and  the
applications considered in the previous sections, the functionals here
are  locally weighted with  the density  $\rho(\x)$. Notice,  that for
inhomogeneous densities  the vector valuations $W_\nu^{0,1}[\rho(\x)]$
in  general  do  not  vanish  for  a convex  body  $K$  and  that  the
functionals  do  inherently  depend  on  the  location  of  $K$.   The
following ansatz  for the excess  free energy functional proves  to be
useful:
\begin{equation}\label{eq:ansatz}
\begin{array}{c} 
 \beta {\f}_{ex}[\rho(\x)] =   \  f_1(W_0^{0,0})~\;W_3^{0,0}+f_2(W_0^{0,0})~\;W_1^{0,0}W_2^{0,0} 
+ f_3(W_0^{0,0})~\;(W_2^{0,0})^3   \cr  
  \cr 
+f_4(W_0^{0,0})~\; W_1^{0,1}  W_2^{0,1}+ 
f_5(W_0^{0,0})~\; W_2^{0,0}  (W_2^{0,1}  W_2^{0,1})\;\;\;,
\end{array} 
\end{equation}
where we drop the argument $[\rho]$ for convenience.  This ansatz is
inspired  by   the  observation  that  the   physical  dimensions  are
$\displaystyle [W_3^{0,0}]=[W_1^{0,q}W_2^{0,q}]=[(W_2^{0,q})^3]=[\beta
f_{ex}]=[{\rm   length}]^{-3}$,     that  $[W_0^{0,0}]=1$   is   a
dimensionless   weighted    density. 
\\
   To   obtain    the   functions
$f_i(W_0^{0,0})$ one may apply an argument from scaled particle theory
for a  homogeneous fluid  \cite{beisbart_reiss:rigid}, namely, that  the excess
chemical  potential for big  spheres of  radius $\sigma/2\to\infty$  is $\lim
\mu/V \to p$  with the volume $V=\frac{ \pi}{6}  \sigma^3$, the chemical
potential  $\mu =  \frac{\partial {\f}_{ex}}{\partial  \rho}$ and the
pressure $p$.  Within
the  present  ansatz  \eqref{eq:ansatz}  one  finds  in  this  limit  the
reversible work to create a cavity of volume $V$,
$\mu = \sum_{\nu} \frac{\partial  {\f}_{ex}}{\partial W_\nu^{0,0}}
\frac{\partial W_\nu^{0,0}}{\partial \rho}$, 
and therefore in the limit $\sigma/2\to\infty$ the chemical potential $\mu/V\to \frac{\partial  {\f}_{ex}}
{\partial W_0^{0,0}}$. 
Using the equation of state
$\Omega = - p V$, i.e.,  $p = - {\f}_{ex} + \sum_{\nu}
\frac{\partial {\f}_{ex}}{\partial W_\nu} W_\nu+ W_3^{0,0}$ 
one gets 
the well known 
 'scaled particle' differential equation
\begin{equation}\label{eq:diffequ} 
\frac{\partial {\f}_{ex}}{\partial W_0^{0,0}} \; = \; -{\f}_{ex} + \sum_{\nu=0}^{3} 
\frac{\partial {\f}_{ex}}{\partial W_\nu^{0,0}}W_\nu^{0,0}\; +\frac{\partial {\f}_{ex}}{\partial W_1^{0,1}}  W_1^{0,1}+ \frac{\partial {\f}_{ex}}{\partial W_2^{0,1}} W_2^{0,1}\; +\;W_3^{0,0} \;. 
\end{equation} 
Within the ansatz \eqref{eq:ansatz} the 
solution of the differential equation reads 
\begin{equation}\label{eq:intrinsicenergy} 
\begin{array}{ll} 
\displaystyle  
\beta f_{ex}[\rho(\x)]  \; \; = \; \; & 
\displaystyle 
  -\frac{3W_3^{0,0}}{  4\pi }   \ln ( 1-W_0^{0,0}  ) 
+ \frac{9}{ 4\pi} \frac{W_1^{0,0}W_2^{0,0} - 
W_1^{0,1} W_2^{0,1}}{ 1-W_0^{0,0}}  \cr 
   &  \cr 
  & \displaystyle  
+ \frac{ 9W_1^{0,0}}{  8\pi} \frac{ (W_1^{0,0})^2 -  3 W_1^{0,1}  W_1^{0,1} 
}{  (1-W_0^{0,0})^2  } \;.  
\end{array} 
\end{equation} 
This functional yields excellent results for thermodynamic quantities
and  for structures  in the  fluid phase  of hard  spherical molecules
\cite{beisbart_roth:depletion}.  For instance,  in Figure~\ref{fig:beisbart_ro}  the inhomogeneous density
profile  $\rho(z)$  close to  a  hard wall  is  shown  as compared  to
simulation     results.  
\\
 Unfortunately,    the     functional    in
Eq. \eqref{eq:intrinsicenergy} was derived  only for hard spheres in three
dimensions;  however, a  proposal  has been  made  for general  convex
bodies  \cite{beisbart_rosenfeld:molecular} based on  an approximative
decomposition of $f_{ex}$ in the low density limit.  Here, we show how
Rosenfeld's    idea    of    decomposing    $f_{ex}$    for    spheres
\cite{beisbart_rosenfeld:free}    may   be    extended    exactly   to
non-spherical convex bodies  by applying integral geometric techniques
such as  kinematic formulae and  an explicit local expression  for the
Euler  characteristic  $\chi(K\cap  K')$  of  two  overlapping  bodies
\cite{beisbart_mecke:fundamental}.  The main idea is to decompose the exact expression for
the local  excess free energy  $f_{ex}[\rho(\x)]$ at low  densities in
terms of the vector-valued Minkowski  functionals. To this end one has
to evaluate the configurational integral  $\int \dd K = \int \dd R\int
\dd\x$  over  all  orientations  $R$  and positions  $\x$  of  a  body
$K=RK_{\x}$, namely,
\begin{equation}\label{eq:rosenfeld1}
\begin{array}{cl} 
  &  \;\;\; 
\displaystyle 
\frac{4\pi }{ 3} \int \dd\x \; \beta f_{ex}[\rho(\x)] \;\; \rightarrow \;\;  
  \frac{2\pi}{ 3} \int \dd K  \int \dd K'  \; \chi(K_{\x} \cap K_{\x\,'})  \;\; \rho(\x) \rho(\x\,') 
 \cr
  & \cr 
 =  \; \;\; \; &\displaystyle\int_{\R^d} \dd\x\; \;    W_3^{0,0}[\rho(\x)] W_0^{0,0}[\rho(\x)] \cr  
   \;\; 
+ \; &\displaystyle \int_{\R^d} \dd\x\;   
3 W_1^{0,0}[\rho(\x)] W_2^{0,0}[\rho(\x)] - 3 W_1^{0,1}[\rho(\x)]  
W_2^{0,1}[\rho(\x)]  \cr
 \;\; + \;    &  \displaystyle \int_{\R^d} \dd\x \;\; \int_{\br} 
\dd K_{\x} 
\rho(\br) \int_{\br\,'}  \dd K'_{\x} \rho(\br\,') 
  \frac{\kappa_1-\kappa_2 }{ 2}   \frac{\left( \bv_1\bv_1 -\bv_2\bv_2 \right)_{ij}  
(\n\,'\n\,')_{ij} }{ 1+\n\n\,'} 
\cr  
\end{array}
\end{equation}
where  $\bv_i$  denotes  the   principal  direction  of  the  curvature
$\kappa_i$  at $\br$  and $\n$  ($\n\,'$)  are the  normal vectors  of
$RK_{\x}$ ($R'K_{\x}$) at $\rb$ ($\rb'$), respectively.  The integral over rotations $R$
and over the  surface of the body $K_{\x}$ centered  at $\x$ is written
as $\int_{\br} \dd K_{\x} = \int \dd R \int_{\br\in \partial RK_{\x}}
\dd S^{d-1}$.    For  spheres   with  $\kappa_1=\kappa_2$    one  recovers
immediately the low-density limit of the intrinsic  free energy given
by Eq.  \eqref{eq:intrinsicenergy}. But  unfortunately the last term on
the  right hand  side  of Eq.   \eqref{eq:rosenfeld1}  vanishes only  for
spheres, so  that only for hard  spheres a finite  number of Minkowski
functionals is,  by accident, sufficient to decompose  the excess free
energy into  contributions stemming alone from  the individual bodies,
whereas for  non-spherical shapes in arbitrary  dimensions an infinite
number of  fundamental curvature measures  is required.  Nevertheless,
based  on the most  general analytical  expression \eqref{eq:rosenfeld1},
several approximations  by a finite number  of tensor-valued Minkowski
functionals (including  the
proposal  in   \cite{beisbart_rosenfeld:molecular}) may be  proposed  and numerically  tested   which   make  Rosenfeld's  density
functional  applicable to  general  convex bodies,  in particular,  to
spherocylinders and ellipsoids \cite{beisbart_mecke:fundamental}.
\section{Conclusions}
\label{sec:beisbart_co}
Concluding,  we argue  that a  new perspective  emerges as  regards the
Minkowski  functionals.   Instead  of   marking  a  small   number  of
distinguished measures (which, of course, they do in a certain sense),
the Minkowski functionals are embedded  into a more extensive class of
Minkowski  valuations.    These  measures  generalize   the  Minkowski
functionals without  leaving the  framework of integral  geometry. The
essential  step to  extend the  scalar functionals  is to  replace the
motion invariance by motion  covariance.  This  step yields a
hierarchy of tensor-valued  measures with  the   Minkowski
functionals at the bottom.  The
higher-rank valuations are moments of the Minkowski functionals where
local  geometry is  weighted with  either the  position or  the normal
vector.  The  full, comprehensive  description  of the  tensor-valued
measures is  still a  mathematical challenge, although  some important
results  concerning  the integral  geometry  of  tensor measures  have
already   been   achieved,   especially   in   low   dimensions   (see
\cite{beisbart_schneider:tensor2}).
\\
The   extensions   of  the   Minkowski   functionals   are  not   only
mathematically  interesting;   rather  they can be motivated on
physical grounds. Since the Minkowski functionals proved to be useful for a
number of applications, we are justified to expect that also their
generalizations may be suitable tools.   Using  physical  applications  at  various
length  scales we  exemplified how  higher-rank  Minkowski valuations
complement the information featured by the scalar functionals. From
this point of view we hope that the higher-rank Minkowski valuations
will become a standard tool for morphometry and some other fields of
research concerned with pattern formation.


\section*{Acknowledgement}
We thank C.  M\"ollen\-hoff  for providing us with recent observations
of  spiral  galaxies.   Furthermore,   we  thank  C.   Levit  for  the
permission to show the images of  the molecules and J. Colberg for the
GIF cluster data. We thankfully acknowledge comments on the manuscript
by M.~Kerscher and R.~Schneider. Last but not  least we thank  the participants of
the second Wuppertal conference for vivid discussions.

\bibliographystyle{springer}\bibliography{my}

\end{document}